\documentclass[aps,prx,twocolumn,superscriptaddress,showpacs]{revtex4-1}
\usepackage{float}
\usepackage{graphicx}  
\usepackage{color}
\usepackage{xcolor}
\usepackage{bm}        
\usepackage{braket}
\usepackage{amssymb}   
\usepackage{epstopdf}
\usepackage{xfrac}
\usepackage{cancel}
\usepackage{soul}
\usepackage{amsmath}
\usepackage{amsthm}
\usepackage{amsfonts}
\usepackage{bbm}
\definecolor{darkblue}{rgb}{0.1,0.2,0.6}
\definecolor{darkred}{rgb}{0.8,0.1,0.2}
\definecolor{darkgreen}{rgb}{0.31,0.62,0.24}
\definecolor{OliveGreen}{cmyk}{0.64, 0, 0.95, 0.40}
\usepackage[colorlinks,citecolor=darkblue,linkcolor=darkblue,urlcolor=darkblue]{hyperref} 
\usepackage{enumerate}
\usepackage{setspace}
\usepackage{url}  
\usepackage{mathrsfs}

\usepackage[percent]{overpic}

\newcommand{\OO}{{\rm O}}
\newcommand{\SO}{{\rm SO}}

\newcommand{\Tr}{\text{Tr}}

\newcommand{\Pb}{\textsc{P}}

\newcommand{\hgamma}{\hat \gamma}
\def\i{\textbf{i}}

\usepackage{algorithmicx}
\usepackage{algorithm}
\usepackage[noend]{algpseudocode}

\newcounter{protocol}
\makeatletter

\begin{document}

\title{Measurement-induced entanglement  
transitions in quantum circuits of non-interacting fermions: Born-rule versus forced measurements}

\author{Chao-Ming Jian}
\affiliation{Department of Physics, Cornell University, Ithaca, New York 14853, USA}

\author{Hassan Shapourian}
\affiliation{Microsoft Station Q, Santa Barbara, California 93106 USA}

\author{Bela Bauer}
\affiliation{Microsoft Station Q, Santa Barbara, California 93106 USA}

\author{Andreas W. W. Ludwig}
\affiliation{Department of Physics, University of California, Santa Barbara, CA 93106, USA}

\date{\today}

\begin{abstract}

We address entanglement transitions in monitored random quantum circuits of non-interacting fermions, in particular, the question of whether Born-rule and forced measurements yield the same universality class. For a generic circuit with no symmetry other than fermion parity, acting on a one-dimensional Majorana chain, we numerically obtain several critical exponents, providing clear evidence that the two transitions with Born-rule and forced measurements are in different universality classes. We provide a theoretical understanding for our numerical results by identifying the underlying statistical mechanics model which follows from the general correspondence, established in Jian \emph{et al.}, Phys.~Rev.~B 106, 134206, between non-unitary circuits of non-interacting fermions and the ten-fold Altland-Zirnbauer (AZ) symmetry classes. The AZ class is the same for Born-rule and forced measurements of the circuits.
For the circuit under consideration (in AZ class DIII), the statistical mechanics model describing the transition is the principal chiral non-linear sigma model whose field variable is an $\SO(n)$ matrix in the replica limits $n\to 0$ and $n\to 1$ for forced and Born-rule measurements, respectively. The former is in an Anderson localization universality class while we show that the latter is in a novel universality class beyond Anderson localization. 
Both entanglement transitions are driven by proliferation of $\mathbb{Z}_2$ topological defects. The different replica limits account for the difference in the universality classes. Furthermore, we provide numerical and symmetry-based arguments that the entanglement transition in the previously-studied monitored circuit of Majorana fermions based on the loop model with crossings, a highly fine-tuned circuit, belongs to a universality class different from both transitions in the generic circuits discussed in this paper.

\end{abstract}

\maketitle

\tableofcontents

\section{Introduction}
The past few years have witnessed 
 an extremely rapid expansion of research activity in the area of
the dynamics of open many-body systems, which provides new insights into the organizing principles of universal collective quantum behavior.
In particular, the study of monitored quantum dynamics has led to the discovery of novel dynamical entanglement phases and measurement-induced entanglement phase transitions between them \cite{LiChenFisher2018MIPT, Skinner2019MIPT, Chan2019MIPT, ChoiBaoQiEhud2020MIPT, GullansHuse2020Purification, Guallans2020ScalableProbes, JianYouVassuerLudwig2020MIPT, BaoChoiAltman2020}. These entanglement phases and phase transitions
occur in
intrinsically {\it non-equilibrium} many-body quantum systems, and thus lie beyond conventional frameworks and paradigms for many-body physics in equilibrium.
The key to their identification is the investigation of their corresponding entanglement dynamics, especially of the 
entanglement entropy (EE) 
scaling at late times, in individual quantum trajectories of the monitored quantum systems. 

A fruitful setting to study 
dynamical entanglement phases and measurement-induced transitions is given by monitored quantum circuits where the quantum many-body systems' evolution is not only driven by unitary 
gates but also ``interrupted" by
events where the system is {\it monitored} or {\it measured} by an observer or an environment, obtaining
a record of measurement outcomes. The evolution/history of the system associated with a specific set of measurement outcomes is called a {\it quantum trajectory}. 
Rich phenomenology in entanglement dynamics has been found in many monitored quantum circuit models \cite{LiChenFisher2019, Szyniszewski2019, Tang2020nonintegrable, Vasseur2020TreeTN, BarkeshiliTopoCircuit2021, SangHsieh2021, Sang2021Negativity, Ippoliti2021MeasurementOnly, FujiAshida2020, LuntPal2020, LangBuchler2020, Vijay2020VolumeLaw, Nahum2021AlltoAll, BaoChoiEhud2021SymmetryEriched, Turkeshi2020TwoD, Buchhold2021Dirac, Vasseur2022ChargeSharpen, Vasseur2022U1, Ippoliti2022SpacetimeDual, 
Pixley2022Multifractality, LiVasseurFisherLudwig2021, LiVasseurFisherLudwig2021, Jian2020, Cao2019FF, NahumSkinnerMajDefect2020FF, ChenLiFisherLucas2020FF, LuGrover2021SpacetimeDualFF, Buchhold2022FF, TangChenZhu2021FF,LavasaniLuoVijay2022, Sriram2022KitaevModel, PotterVasseur2021EntanglementDynamics, FisherKhemaniNahumVijay2022RQC,DiehlPRL2021FF}.
Despite the ubiquity of measurement-induced entanglement transitions in these 
systems, 
knowledge of the nature of these transitions, particularly their universality classes, is rather limited. 
Many previous studies focused, for the most part, on circuits of qubits or qudits. Beyond certain limits  such as that of infinite on-site Hilbert space  dimension~\cite{JianYouVassuerLudwig2020MIPT, BaoChoiAltman2020}, and other ways the same universality class can be formulated in the circuit context \cite{Skinner2019MIPT,NahumSkinnerMajDefect2020FF,SangHsieh2021, BarkeshiliTopoCircuit2021}, systematic and controlled analytical studies of entanglement phase transitions in  quantum circuits of qubits (or qudits)
have been a major challenge. 

The present paper addresses 
measurement-induced entanglement transitions in monitored quantum dynamics of non-interacting fermions. In such dynamics, both the unitary evolution and the measurements preserve the ``Gaussianity" of the non-interacting fermionic states in each quantum trajectory. (More details are formulated in 
quantum circuit language in Sec. \ref{LabelSectionMonitoredGaussianCircuitsGeneral}). In general, the types of entanglement phases realized in non-interacting fermion systems will be different from those in systems of qudits.
Nevertheless, measurement-induced entanglement transitions are also found to be a generic phenomenon in many examples of non-interacting fermion 
circuits and incarnations of them formulated in the language of other degrees of freedom
that have been discussed in the literature \cite{Jian2020,Cao2019FF,NahumSkinnerMajDefect2020FF,ChenLiFisherLucas2020FF,DiehlPRL2021FF,LuGrover2021SpacetimeDualFF,Buchhold2021Dirac,Buchhold2022FF,TangChenZhu2021FF,LavasaniLuoVijay2022,Sriram2022KitaevModel}.
It is natural to expect that non-interacting fermionic 
circuits are more tractable than those of qubits (qudits).
Hence, they present us with an opportunity to acquire a deeper insight into the underlying mechanisms and the universality classes of the entanglement transitions. Furthermore, the knowledge acquired from studies
of non-interacting fermionic circuits will lay the foundation for the study of the quantum circuits of qudits, which can be conceptually viewed as interacting versions of quantum circuits of fermions.

Circuits of non-interacting fermions are also commonly referred to as fermionic {\it Gaussian} circuits. It was shown in recent work \cite{Jian2020}
that all fermionic Gaussian circuits, whether monitored by measurements, or, in general, subject to a non-unitary time evolution, are described and classified in terms of a common, unifying framework. This unifying framework is based on a general correspondence established in Ref.~\onlinecite{Jian2020} between fermionic Gaussian circuits acting on systems in $d$ spatial dimensions and systems of non-interacting fermions in $(d+1)$-dimensional space subject to static (Hermitian) Hamiltonians, or equivalently, subject to unitary evolutions.
This correspondence applies to any fixed realization of the circuit, with or without translational invariance in space or time.
Owing to this correspondence, any fermionic Gaussian circuit can be classified, via its corresponding static Hamiltonian system, 
according to the Altland-Zirnbauer (AZ) ten-fold symmetry classification~\cite{AltlandZirnbauerPRB1997,RyuSchnyderFurusakiLudwigNJPhys2010}. 
In the monitored Gaussian circuit, each collection of measurement outcomes from the entire space-time history
of the circuit evolution labels a particular quantum trajectory corresponding to
a specific circuit realization. Studying the behavior of a monitored Gaussian circuit in $d$ spatial dimensions averaged over
all quantum trajectories is equivalent to the 
study of the behavior of the corresponding Hamiltonian problem
of non-interacting fermions,
averaged over static ``disorder'' in the space on which the Hamiltonian acts, which is 
the $(d+1)$-dimensional space-time of circuit.
Thus, the correspondence established in Ref.~\onlinecite{Jian2020} links the area of monitored fermionic Gaussian circuits (even more generally, of random non-unitary Gaussian circuits) and the classic area of Anderson localization by offering a framework that encompasses both. 

Yet, it is important to stress that, as we will show in the present paper,
entanglement transitions occurring in the monitored Gaussian
circuits include novel universality classes beyond
Anderson localization transitions, even though both can be described within the same framework and follow the same AZ symmetry classification. The circuit context is a source of new universality classes and novel
physics that do not exist for Anderson localization transitions. As we will see below, a case in point is the system discussed in this paper.

Monitored Gaussian circuits can
give rise to novel
universality classes beyond Anderson localization physics for the following reasons. Within a given symmetry class, there are at least two physically different types of measurements (at the level of how the statistical weights are assigned to each quantum trajectory): {\it Born-rule} measurements and {\it forced} measurements. The former 
weighs the quantum trajectories according to 
its classic Born-rule probability
associated with the measurements, while the latter ``forces" an equal-weight distribution on all the quantum trajectories (more details in Sec. \ref{sec:Born-rule_vs_Forced}). The result of Ref. \onlinecite{Jian2020} directly implies that the monitored Gaussian circuits with forced measurements are equivalent to Anderson 
localization problems. Therefore, the forced-measurement-induced entanglement transitions are identical to Anderson localization transitions in the corresponding AZ symmetry classes.
Ref.~\onlinecite{Jian2020} also implies that forced-measurement transitions are
identical to the corresponding transitions in Gaussian random tensor networks in the same  AZ symmetry classes.
Note that a forced-measurement protocol - which can be used 
as a reformulation \cite{Jian2020} of 
the  Gaussian random tensor network - can only be implemented using post-selection and thus requires exponential overhead in the measurement outcomes.
 The ability to view monitored circuits with forced measurements as a reformulation of random tensor networks also extends, beyond the non-interacting fermion setting,
to circuits of qudits~\cite{Nahum2021AlltoAll}.
While such a reformulation is not necessary, it will prove conceptually convenient  
for making  a comparison  with the Born-rule case most direct~\footnote{As an aside, we note that the term ``forced measurement'' has been used with a different meaning in the
literature of measurement-based topological quantum computation, where a measurement outcome is forced using a ``repeat-until-success'' approach implemented by a probabilistically-determined
adaptive series of operations.
~\cite{BondersonFreedmanNayakMeasurementOnlyPRL2008,BondersonFreedmanNayakMeasurementOnlyAnnPhys2009,TranBocharovBauerBondersonSciPost2020}}.
An important question is whether in a given AZ symmetry class for the Gaussian circuits,
an entanglement transition induced by Born-rule measurements shares the same
universality class as the 
forced-measurement/random tensor network one.
In the context of monitored quantum circuits of qubits (qudits) 
the same question was raised in Ref. \onlinecite{Nahum2021AlltoAll},
building on the results of 
Refs.~\onlinecite{VasseurPotterYouLudwigRTNPRB2019,JianYouVassuerLudwig2020MIPT, BaoChoiAltman2020},
and a difference between universality classes associated with
the two types of measurements was conjectured. On the other hand,
for monitored Clifford circuits
(``stabilizer circuits"),  Born-rule and forced measurements yield the same universality class of the entanglement transition \cite{LiVasseurFisherLudwig2021}.
The present paper provides both concrete numerical evidence and analytical arguments that demonstrate the difference between entanglement transitions induced by Born-rule measurements and forced measurements in monitored Gaussian circuits. 
In particular, we show that
the monitored Gaussian circuits with Born-rule measurements give rise to novel universality classes beyond Anderson localization transitions. 

In this paper, we answer the important question raised above by studying a monitored fermionic Gaussian circuit acting on a one-dimensional Majorana chain. We require the circuit to preserve no symmetry other than the global fermion parity. 
As was shown in Ref.~\cite{Jian2020}, such a fermionic Gaussian circuit belongs to class DIII within the AZ symmetry classification.
We numerically study the phase diagrams of this monitored fermionic Gaussian circuit with both Born-rule and forced measurements. We find that both types of measurements can lead to their
corresponding
measurement-induced entanglement transitions, and that these are in different universality classes. 
To show this, we consider the (square of the) Majorana fermion correlation function at the final time slice of the circuit in the long-time limit. This correlation function turns out to exhibit rich scaling behavior at both the entanglement transitions with Born-rule measurements and that with forced measurements.
 At both transitions the
exponent for the spatial decay of the averaged correlation function significantly differs from that of the typical correlation function. This is a signature of 
so-called  {\it multifractal} scaling
of the correlation function 
\cite{LudwigHierarchies1990,Pixley2022Multifractality}
(briefly reviewed in Appendix \ref{LabelSectionAppendixMultifractal}).
Specifically, we show that the entanglement transitions with Born-rule and forced measurements exhibit widely different decay exponents of the {\it typical} correlation function.
Moreover, while the logarithm of the correlation function is self-averaging (giving rise to the typical scaling exponent),
the statistical fluctuations about its average define another universal quantity which differs widely between the transitions with Born-rule and forced measurements. (See
Table \ref{Tab:exponent_summary} for a summary.)
This provides strong
evidence that the two transitions belong to different universality classes.

Furthermore, we identify the statistical-mechanics models describing these 
monitored Gaussian circuits with Born-rule and forced measurements. These models provide a theoretical understanding of the numerically observed different universal critical behavior.
This identification follows from
the correspondence established in Ref.~\onlinecite{Jian2020}. The key point is that,
as already mentioned above, both monitored Gaussian circuits, those with Born-rule and those with forced measurements, are described by the same
AZ symmetry class which is class DIII for the circuits under consideration. This common AZ symmetry class dictates that the entanglement transitions in both types of circuits is described 
(in continuum language) by a two-dimensional principal chiral non-linear sigma model (NLSM) whose field variable is a group element in the 
special orthogonal group $\SO(n)$
(the target space), where $n$ is a replica index. 
This NLSM possesses an $\SO(n) \times \SO(n)$ symmetry.
The different replica limits $n\to 0$ and $n\to 1$  correspond to the monitored Gaussian circuits with forced and with Born-rule measurements, respectively. The difference in replica limits is the origin of the different critical exponents observed numerically. It is crucial that in both limits, the number $n(n-1)/2$ of degrees of freedom (equal to the dimension of the group manifold $\SO(n)$)  goes to zero so as to yield
a constant partition function (and a vanishing conformal central charge) in the limit. Both entanglement transitions are driven by proliferation 
of $\mathbb{Z}_2$ topological defects. These transitions parallel the two-dimensional disorder-driven metal-insulator
transition studied in Ref.~\onlinecite{FuKane2012Sympletic} which is in a different symmetry class.
 We stress again that for Born-rule measurements, the transition is {\it not} an Anderson localization transition but instead a novel transition unique to the physics of monitored random circuits.

We close by comparing the two monitored Gaussian circuits discussed in the present paper with the previously-studied circuit based on the loop model with crossings \cite{NahumSkinnerMajDefect2020FF},
which is a highly fine-tuned version of the monitored Gaussian circuit of Majorana fermions. We provide numerical and
analytical evidence showing that the entanglement transition in this loop-model-based circuit
is in a
universality class different from that of
both generic entanglement transitions discussed in the present paper. Regarding the numerical evidence, we observe that the numerical values for the correlation length exponent $\nu$ are significantly different. Regarding the analytical
evidence we 
note that the statistical-mechanics model of the loop-model-based circuit
has a different symmetry, which is $\SO(n)$ in the limit $n\to 1$, different from that of the statistical-mechanics model of the circuits discussed in the present paper, which is $\SO(n)\times \SO(n)$ in the limits $n\to 0$ and $n\to 1$, for forced and Born-rule measurements, respectively. Given the different symmetries, one does not expect a relationship between the universality classes of the transitions in these two types of Gaussian circuits.

The rest of the paper is organized as follows: In Sec.~\ref{LabelSectionMonitoredGaussianCircuits}, 
we introduce the general construction of monitored Gaussian fermionic circuits, explain the two measurement schemes, and how we parameterize our quantum circuit. In Sec.~\ref{sec:numerics}, we present our numerical results including the phase diagram and several critical exponents for the entanglement phase transition
for each of the two measurement schemes. In Sec.~\ref{sec:replica_theory} 
we introduce the statistical mechanics model which provides a theoretical understanding of our numerical results, and
also make comparisons with the loop model with crossings. We finish our paper with several concluding remarks and possible future directions in Sec.~\ref{sec:conclusions}. In four appendices, we further provide details of our numerical simulations (including some additional data) and a discussion on 
the multifractal scaling of correlation functions which is a key tool we use in our analysis.

\section{Monitored Gaussian Circuit Dynamics of Majorana Chain}
\label{LabelSectionMonitoredGaussianCircuits}

\subsection{General construction}
\label{LabelSectionMonitoredGaussianCircuitsGeneral}

We study the monitored Gaussian circuit dynamics of a chain of Majorana fermion modes. Denote the Majorana mode on the $i$th site 
by $\hgamma_i$. It turns out that a class of monitored Gaussian circuits can be built from two-site unitary gates of the form $e^{ \i \alpha  \hat \gamma_i \hat \gamma_{i+1}}$ (with some purely-imaginary-valued parameter $\alpha$) 
and projective fermion-parity measurements on pairs of neighboring sites. 
More generally, monitored Gaussian circuits are defined by the property that if the initial state is a Gaussian state of non-interacting fermions, the state remains Gaussian under the circuit evolution for any given set of measurement outcomes, namely the Gaussianity of the state is maintained along each quantum trajectory. 
Gaussian fermionic states are the most general states that obey Wick's theorem, i.e., all their equal-time correlation functions are fully characterized by their equal-time two-point correlation 
function \cite{Bravyi2004FermionicOptics}.
The most general form of monitored Gaussian circuits can be formulated using the language of {\it generalized measurements}, which we review below. For this paper, we aim to study the universal properties of the entanglement dynamics and entanglement phase transitions in monitored Gaussian-circuit dynamics of a one-dimensional Majorana chain. Such universal behavior is expected to be independent of the specific realizations of the monitored Gaussian circuit. We will see below that formulating the monitored Gaussian circuit using generalized measurement gives us an advantage when we compare the monitored Gaussian circuits respecting the Born-rule and the circuits with forced measurements. 

Let us first briefly review the formalism for generalized measurements. A generalized measurement is defined by an ensemble of Kraus operators $\mathcal{M} = \{K_m\}$ where $m$ labels the possible measurement outcomes and where
the ensemble of operators $K_m^\dag K_m$ forms a positive operator-valued 
measure (POVM), i.e. $\sum_m w_m K_m^{\dag} K_m = \openone$ with a non-negative weight $w_m\in \mathbb{R}_{0+}$ for each $m$.
Under a generalized measurement, each measurement outcome, labeled by $m$, corresponds to a quantum trajectory, namely the evolution of an incoming state $|\psi\rangle$ to $\frac{K_m |\psi \rangle}{ \| K_m |\psi \rangle \|}$ when the $m$th measurement outcome occurs. This quantum trajectory occurs with the Born-rule probability $p_m = w_m \langle \psi| K_m^\dag K_m |\psi \rangle$. The POVM condition ensures that the 
probability is normalized, $\sum_m p_m = 1$,
for any incoming state $|\psi \rangle$. 
As an example, for the projective measurement of the fermion parity $\i \hgamma_1 \hgamma_2$ associated with two Majorana modes $\hgamma_{1,2}$, the ensemble of Kraus operators is given by 
$\{K_m\}=\left\{\frac{1+m \i \hgamma_1 \hgamma_2}{2} \right\}_{m = \pm}$ ,
and $w_{m=\pm}=1$.
Another example of generalized measurements is given by the probabilistic projective measurement: A projective measurement of $\i \hgamma_1 \hgamma_2$ is implemented with classical probability $p$, and no measurements occur with probability $1-p$. In this case, the Kraus-operator ensemble is given by 
$\{ K_m \}= \left\{\sqrt{1-p}, \sqrt{p}\frac{1+ \i \hgamma_1 \hgamma_2}{2},  \sqrt{p}\frac{1 - \i \hgamma_1 \hgamma_2}{2} \right\}$.
When the number of possible measurement outcomes is finite, one can absorb the weight $w_m$ into the definition of the Kraus operator $K_m$ by re-scaling $K_m \rightarrow \sqrt{w_m} K_m$.
When measurement outcomes form a continuous set,
$w_m$ is interpreted as the measure for the integration over this space of outcomes.
Conceptually, a generalized measurement $\mathcal{M} = \{K_m\}$  can be viewed as a representation of the combined effect of a sequence of unitary rotations and projective measurements \footnote{The more detailed statement 
is that any generalized measurements can be implemented by combining unitary operations and projective measurements (possibly with the help of ancillary degrees of freedom).
See for example Ref. \onlinecite{watrous2018theory} (in particular Section 2.3 and Theorem 2.42).}.
One expects that the universal behavior of the monitored circuit, especially the universality classes of the measurement-induced entanglement transitions, should be insensitive to whether one uses
(probabilistic) projective measurements (together with unitary gates) or more generic generalized measurements in the circuit.

\begin{figure}
\centering
   \includegraphics[scale=0.65]{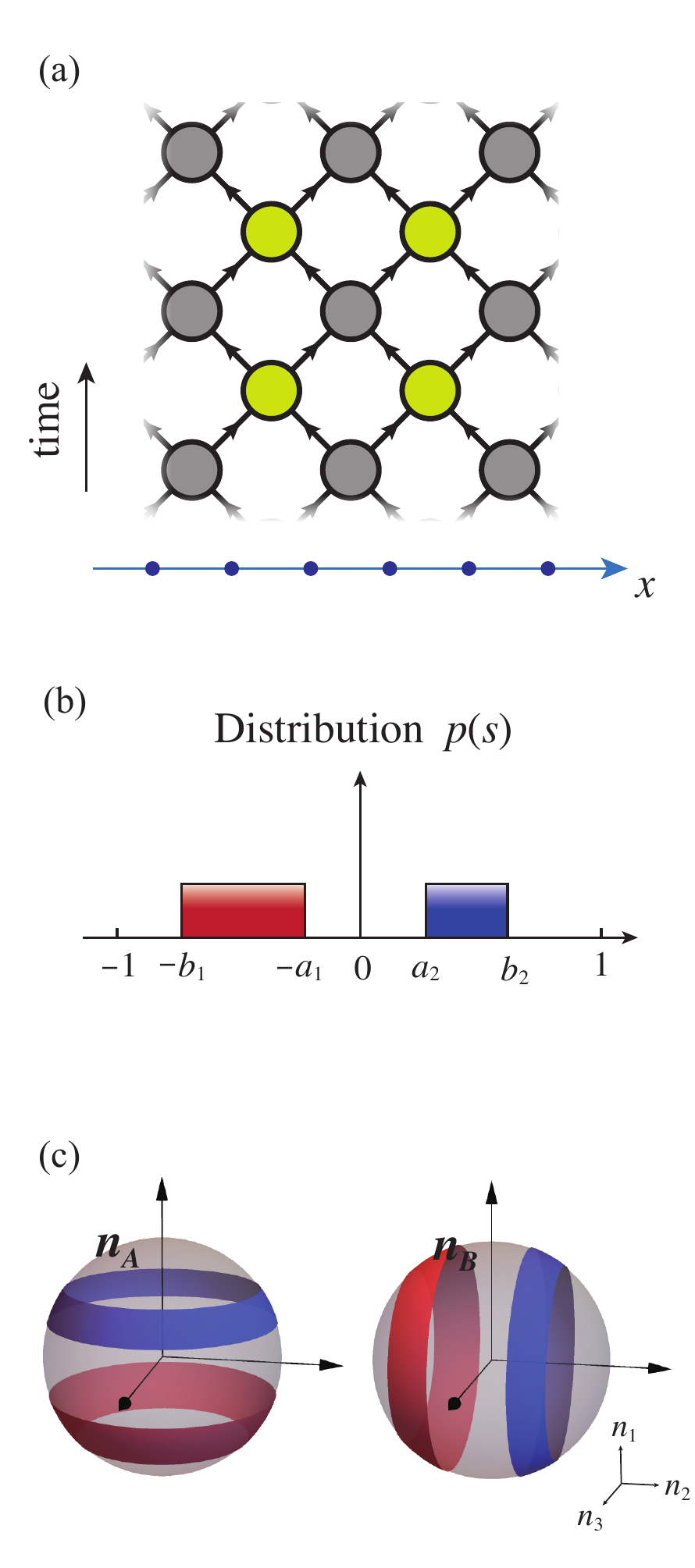}
   \caption{The spacetime geometry of the monitored Gaussian circuit acting on a Majorana chain (along the $x$ axis) is depicted in (a). The gray and yellow gates, representing two types of generalized measurements with different Kraus-operator ensembles, occupy the $A$ and $B$ sublattices of the circuit geometry, respectively. An ingredient of the definition of these Kraus-operator ensembles is the variable $s=\cos\theta$ which follows the distribution $p(s)$ introduced in Eq. \eqref{eq:p-dist} and shown in (b). The visualization of the ensembles of the unit vectors $n_A$ and $n_B$ is provided in (c).
   }
   \label{fig:diii-circuit}
\end{figure}

As it follows from the definition of Gaussian circuits, the measurements are required to preserve the Gaussianity of the state. Therefore, in a system consisting  of Majorana fermion modes, the most general form of corresponding Kraus operators is given by
$\exp\left(\sum_{ij} \i  c_{ij} \hgamma_i \hgamma_j \right)$ with 
complex coefficients $c_{ij} \in \mathbb{C}$. For our study, we consider monitored Gaussian circuits constructed from local gates (including unitary operations and measurements) acting on a one-dimensional Majorana chain. It suffices to focus on the gates that only act on nearest-neighbor pairs of Majorana modes. The corresponding Gaussian circuit has a geometry depicted in Fig. \ref{fig:diii-circuit} (a). Each nearest-neighbor two-site gate, 
depicted as a gray or a yellow disk in Fig. \ref{fig:diii-circuit} (a), is {\it independently} drawn from a Kraus-operator ensemble $\{K(\vec{n})\}$ with 
\begin{align}
\label{LabelEqKvecn}
K(\vec{n}) = (1-n_1^2)^{\frac{1}{4}} e^{ -\i  \alpha(\vec{n})  \hat \gamma_i \hat \gamma_{i+1}}  
\end{align}
with the (generally complex)
coefficient $\alpha(\vec{n})$ parameterized by a three-component {\it unit} vector $\vec{n}=(n_1, n_2, n_3)$:
\begin{align}
\label{LabelEqProbps}
e^{2 {\rm Re} (\alpha)} = \left(\frac{1+n_1}{1-n_1}\right)^{\frac{1}{2}},~~
e^{2 \i \, {\rm Im} (\alpha)} = \frac{n_2 - \i n_3}{(n_2^2 +n_3^2)^{\frac{1}{2}}}.
\end{align}
The unit vector $\vec{n}$ physically labels the measurement outcomes of the generalized measurement defined by the ensemble $\{K(\vec{n})\}$.
The Kraus operator $K(\vec{n})$ can be decomposed into a unitary 
operation $e^{ \, {\rm Im}(\alpha(\vec{n})) \, \hat \gamma_i \hat \gamma_{i+1}}$ and a positive-semidefinite Hermitian operation $e^{-\i \, {\rm Re}(\alpha(\vec{n})) \, \hat \gamma_i \hat \gamma_{i+1}}$. The latter implements a weak measurement 
of the fermion parity $\i \hat \gamma_i \hat \gamma_{i+1}$, which is a softened version of the projective measurement that yields a projection 
onto the eigenstates of $\i \hat \gamma_i \hat \gamma_{i+1}$. The sign of $n_1$ determines whether the corresponding quantum trajectory is biased towards the eigenstate of $\i \hat \gamma_i \hat \gamma_{i+1}$ with eigenvalue $-1$ or $1$. The magnitude of $n_1 \in [-1,1]$ determines the strength of this bias. In the case of $n_1 = \pm 1$, the Kraus operator essentially implements the projection onto the eigenstate with $\i \hat \gamma_i \hat \gamma_{i+1} = \mp 1$. Therefore, the probabilistic projective measurement of $\i \hat \gamma_i \hat \gamma_{i+1}$ discussed in the previous paragraph 
can be formulated in terms of the Kraus operator parametrization of Eq.~\eqref{LabelEqKvecn}
upon setting the vector $\Vec{n}$ to three possible values: $\Vec{n}= (\pm 1,0,0)$  and $\Vec{n}= (0,1,0)$. "
Here, it may seem ad hoc to use the unit vector $\vec{n}$ to parameterize the Kraus operator $K(\vec{n})$. We use it because Ref. \onlinecite{Jian2020} has 
shown that 
the most general Gaussianity-preserving gate admits a parametrization using symmetric spaces. When restricted to gates that act on only two Majorana modes, the
corresponding
symmetric space is reduced to the unit sphere $S^2$.

The Kraus-operator ensemble $\{K(\vec{n})\}$ can be described by an ensemble $\mathcal{E}$ of the unit vectors $\vec{n}$ corresponding to each $K(\vec{n})$ and the weight $w(\vec{n})$ that provides a measure on $\mathcal{E}$. 
The POVM condition is written as
\begin{align}
    \openone & = \int_{\vec{n}\in \mathcal{E}} d\vec{n} ~ w(\vec{n}) K^\dag(\vec{n}) K(\vec{n}) 
    \nonumber \\
    & = \int_{\vec{n}\in \mathcal{E}} d\vec{n} ~ w(\vec{n}) (\openone - n_1 \i \hat \gamma_i \hat \gamma_{i+1} )
    \label{eq:POVM_nvec}
\end{align}
which follows from 
Eqs.~\eqref{LabelEqKvecn} and \eqref{LabelEqProbps}.
In principle, one can choose a different ensemble $\mathcal{E}$ for every gate. That means one can perform a different generalized measurement for every pair of neighboring sites at every time step. In this work, we will consider two different ensembles, one for the gray gates (acting on the pair of neighboring sites $(2i,2i+1)$) and the other for the yellow gates (acting on the pair of neighboring sites $(2i-1,2i)$)
shown in Fig. \ref{fig:diii-circuit} (a). Viewing the circuit geometry as a square lattice in spacetime, the gray and yellow gates 
occupy the two different sublattices $A$ and $B$. The details of these ensembles will be given after the current general discussion.

Once the Kraus-operator ensemble is determined for every gate, we obtain an ensemble of Gaussian circuits following the circuit geometry in Fig. \ref{fig:diii-circuit} (a). Each realization of the Gaussian circuit corresponds to a quantum trajectory of the system labeled by a collection of the outcomes 
$\vec{n}$, independently chosen
for every generalized measurement. 
We are interested in the averaged behavior of the monitored Gaussian circuit. That is to say, any physical 
quantity will be averaged over all circuit realizations (and, equivalently, all quantum trajectories). Interestingly, with the same Kraus operator ensembles, different types of weighted averages give rise to different versions of the monitored Gaussian circuit. In particular, for the same Kraus operator ensemble, the Born-rule measurement, and the forced measurements implement two different commonly discussed statistical weights for the average over quantum trajectories, which we will explain in the next subsection.

We stress again that, as established in Ref.~\onlinecite{Jian2020},
there is a general correspondence between fermionic Gaussian circuits and
static Hamiltonian systems of non-interacting fermions.
Via the corresponding static Hamiltonian system, fermionic Gaussian circuits can be classified according to the AZ ten-fold symmetry classification.
The symmetry class 
turns out to play an essential role in determining the universal behavior of the Gaussian circuit. 
Generic monitored Gaussian circuits acting on a chain of Majorana fermion modes
belong to symmetry class DIII within the AZ symmetry classification~\footnote{The symmetries of the AZ classification refer to that of the system with a static Hamiltonian in a spatial dimension
equal to the space-time dimension of the circuit,  which appears in the aforementioned correspondence. Due to the correspondence, the classification of the systems with static Hamiltonian is inherited by the circuits, but the symmetry is generally not.  The monitored Gaussian circuits in this paper respect no symmetry other than fermion parity.}.
It is important to note that the AZ symmetry class is
the same for monitored Gaussian circuits with Born-rule
measurements and those with forced measurements, the
two types of measurements discussed in detail in the following section. Hence, both types of monitored Gaussian circuits under study in this paper are described by symmetry class DIII.

\subsection{Born-rule measurements versus forced measurements}
\label{sec:Born-rule_vs_Forced}
As stated earlier, a generalized measurement is defined by a Kraus operator ensemble $\cal M$ with each Kraus operator corresponding to a measurement outcome, or a quantum trajectory at this measurement. A quantum trajectory for the entire monitored quantum circuit is labeled by a specific collection of outcomes, one from each measurement in the circuit. Each quantum trajectory gives rise to a specific realization of the circuit. The universal behavior of the monitored Gaussian circuit should be characterized by the average behavior of all the quantum trajectories. This averaging depends on the statistical weight of each quantum trajectory. When the Kraus operator ensemble of each measurement is fixed, there are two natural and commonly discussed statistical weights. Generalized measurements with their quantum trajectories averaged according to these two types of statistical weights are referred to as the {\it Born-rule} measurements and the {\it forced} measurements. This work aims to understand the difference between the universal behaviors, the entanglement transitions in particular, of monitored Gaussian circuits with Born-rule measurements and those with forced measurements.

First, we
discuss the statistical weights of quantum trajectories associated with Born-rule measurements. In the context of monitored Gaussian circuits, given the Kraus operator ensemble $\{K(\vec n)\}$ of generalized measurements, the standard Born-rule probability (or probability density) for observing the outcome $\vec{n}$ is given by $w(\vec{n})\langle\psi| K^\dag(\vec{n}) K(\vec{n}) |\psi \rangle$ which depends on the (normalized) state $|\psi \rangle$ of the system prior to the measurement. With Born-rule measurements, the behavior of the monitored Gaussian circuit is given by averaging over all quantum trajectories weighted by their corresponding Born-rule probability/probability density. 

Second, in the
monitored Gaussian circuit with forced measurements, we ``force" each quantum trajectory to appear with equal probability. That is to say, the behavior of the monitored Gaussian circuit with forced measurements is obtained from averaging over all quantum trajectories with equal statistical weight. Even though the natural probability distribution of measurement outcomes in quantum mechanics follows the Born rule, an equal probability distribution of quantum trajectories can be achieved by post-selection and ``re-weighting" each quantum trajectory when we study their statistical average.

The correspondence established in Ref. \onlinecite{Jian2020} implies that monitored Gaussian circuits with forced measurements are equivalent to the disordered non-interacting Hamiltonian system of fermions in one higher spatial dimension and in the corresponding AZ symmetry class. Therefore, the forced-measurement-induced entanglement transition in monitored Gaussian circuits shares the same universality class as the Anderson localization transition in the corresponding AZ symmetry class. Interestingly, as we will explain,
the Born-rule-measurement-induced entanglement transition in monitored Gaussian circuits gives rise to new universality classes beyond Anderson localization transitions.

As we show in the remainder 
of the paper, in the example of AZ class discussed here,
even with the same Kraus operator ensemble in the monitored Gaussian circuit, Born-rule measurements and forced measurements 
can lead to distinct universal behavior.
The specific model this paper focuses on is introduced in Sec. \ref{LabelSubsectionCircuitModel}.
The numerical evidence for the difference between the universality classes of the Born-rule-measurement-induced and the forced-measurement-induced entanglement transitions is presented in 
Sec. \ref{sec:numerics}.
From an analytical perspective, the two types of measurements are associated with different replica limits in a 
replica-based treatment of 
the averaging over quantum trajectories
, which we will elaborate on in Sec. \ref{sec:replica_theory}.

\subsection{Circuit model}
\label{LabelSubsectionCircuitModel}
Now we describe the details of the generalized measurements for the monitored Gaussian circuit that we will study numerically. In the circuit geometry shown in Fig.\ref{fig:diii-circuit} (a), the gray gates (acting on the pairs of neighboring sites $(2i,2i+1)$) and the yellow gates (acting on the pairs of neighboring sites $(2i-1,2i)$) are drawn from two different ensembles associated with two types 
of parametrizations of their respective unit vectors $\vec{n}_A$ and $\vec{n}_B$ in terms of polar angles: 
\begin{eqnarray}
\vec{n}_A &=& (\cos\theta,\sin\theta\cos\varphi,\sin\theta\sin\varphi) \nonumber \\
\vec{n}_B &=& (\sin\theta\sin\varphi,\cos\theta,\sin\theta\cos\varphi). \label{eq:nAB_reparameterized}
\end{eqnarray}
Here, $\theta\in [0, \pi]$ and $\varphi \in [0, 2\pi)$ are random variables chosen independently for each gate (including both the gray and yellow gates). The difference in the parameterization of $\vec{n}_A$ and $\vec{n}_B$ leads to a staggered pattern in the space-time geometry of the circuit. 
The ensemble for $\varphi$ is taken to be the interval $[0,2\pi)$ with a uniform distribution. Let us define the random variable $s = \cos \theta$ where
 $s$ is drawn from the ensemble given by the union of two intervals $[-b_1, -a_1] \cup [a_2, b_2]$ with a uniform distribution $p(s)$ as shown in Fig.~\ref{fig:diii-circuit} (b):
\begin{align}
\label{eq:p-dist}
p(s)=
    \left\{
    \begin{array}{cc}
    \frac{1}{ (b_2-a_2)+ (b_1-a_1)} & \ \  s\in [-b_1, -a_1] \cup [a_2, b_2] \\
    0     & \text{other $s$.}
    \end{array}\right.
\end{align}
The parameters $a_{1,2}$ and $b_{1,2}$ satisfying $0\leq a_1<b_1\leq1$ and $0 \leq a_2<b_2\leq 1$ are the tuning parameters of the monitored Gaussian circuit model we focus 
on, corresponding to a variable staggering in the circuit. 
Phase diagrams of the circuit as a function of these parameters will be obtained in the following sections.  Note that value of $\sin\theta$ is given by $\sin\theta = \sqrt{1-s^2}$ since $\theta\in [0, \pi]$.

{\it Born-rule measurements}:
To ensure the POVM condition Eq. \eqref{eq:POVM_nvec} for each generalized measurement, we need to introduce the 
weights $w_{A/B}(s) = p(s) \Tilde{w}_{A/B}(s)$ for the grey and yellow gates respectively:
\begin{align}
\label{LabelEqWeightsBornRule}
&\Tilde{w}_A(s)=
    \left\{
    \begin{array}{cc}
    \frac{(b_2+a_2)}{(b_1-a_1)}  \frac{(b_1-a_1+ b_2-a_2)}{(a_1+b_1+a_2+b_2)}   & \ \ -b_1\leq s \leq -a_1, \\
    \frac{(b_1+a_1)}{(b_2-a_2)}  \frac{(b_1-a_1+ b_2-a_2)}{(a_1+b_1+a_2+b_2)}   & \ \ a_2 \leq s \leq b_2,
    \end{array}\right. \nonumber
\\
& \Tilde{w}_B(s) = 1 .
\end{align} 
In terms of the new parameters 
$\varphi$ and $s$,
the weighted average over the quantum trajectories in the Gaussian circuit with Born-rule measurements is formally carried out by the integration 
\begin{align}
    & \int d\vec{n}_Y ~ w(\vec{n}_Y)\langle K^\dag(\vec{n}_Y) K(\vec{n}_Y) \rangle 
     \nonumber \\
     &  = \int_{0}^{2\pi} 
    \frac{d\varphi}{2\pi } \int_{-1}^1 ds ~ p(s) \Tilde{w}_Y(s) \langle K^\dag(\vec{n}_Y) K(\vec{n}_Y)\rangle 
    \label{Eq:Born-rule_average}
\end{align}
for every measurement with $Y=A$ or $B$ depending on the sublattice the measurement belongs to in the circuit geometry. Here, the factor $\langle K^\dag(\vec{n}_{Y}) K(\vec{n}_{Y}) \rangle$ is evaluated on the (normalized) state of the system undergoing the corresponding measurements. 
Recall that the difference between the respective Kraus-operator ensembles for the gray and the yellow gates, located on the $A$ and the $B$ sublattice of the square lattice, respectively, comes from the difference between how 
$\vec{n}_A$ and $\vec{n}_B$
are parameterized by $s = \cos\theta$ and $\varphi$ via Eqs. \eqref{eq:nAB_reparameterized}. Thus, tuning the parameters $a_{1,2}$ and $b_{1,2}$ introduced above amounts to a variable staggering of the circuit.

{\it Forced measurements:}
In contrast, for the forced-measurement counterpart of the monitored Gaussian circuit, each quantum trajectory is weighted equally and independent from the state of the system. Therefore, averaging over all quantum trajectories amounts to performing an integral 
\begin{align}
\label{LabelEqForcedMeasurementWeight}
    \int_{0}^{2\pi} 
    \frac{d\varphi}{2\pi } \int_{-1}^1 ds ~ p(s)
\end{align}
for each measurement for both sublattices in the
circuit geometry.

Conceptually, one of the most fundamental differences between the Born-rule measurements and the forced measurements is that the former is associated with a probability distribution of quantum trajectories depending on the state of the system undergoing the measurement while the latter enforces a {\it pre-determined} uniform distribution across all trajectories. As we explain Sec. \ref{sec:replica_theory}, this difference leads to distinct replica limits in the statistical-mechanics description of the circuit. Deforming the detailed form of $p(s)$ is expected not to affect the universal behavior of the circuit.

Here, we would like to comment on the advantage of considering generalized measurements defined in this subsection (defined through Eqs.~\eqref{LabelEqKvecn}, \eqref{LabelEqProbps}). With forced measurements, a general implicit assumption is that all the quantum trajectories are associated with non-vanishing wavefunctions of the system. A subtlety with projective measurements is that, on rare occasions, certain quantum trajectories can have a vanishing wave function, and thus these quantum trajectories appear with zero statistical weight in the forced measurement ensemble, as opposed to the pre-determined weight which defines the forced measurement protocol. However, these rare events are not expected to affect the universal averaged behavior of the circuit with forced measurements. With generalized measurements of the kind introduced in this section, we can completely avoid this subtlety. That is because, unlike the projection operators appearing in the case of projective measurements, none of the Kraus operators we introduced above for the generalized measurements (i.e. those defined in this subsection using Eqs.~\eqref{LabelEqKvecn}, \eqref{LabelEqProbps})
annihilate any possible state of the system. Again, we stress the circuits with generalized (Born-rule or forced) measurements are expected to produce the same universal behavior as those with 
corresponding projective measurements.

\section{Numerical Results}
\label{sec:numerics}

\label{LabelSectionNumericalResults}

\subsection{Born-rule measurements}

\label{LabelSubsectionBornRuleMeasurements}

We numerically simulate the monitored Gaussian circuit with the geometry shown in Fig. \ref{fig:diii-circuit} (a). The Kraus-operator ensembles for the gray and yellow gates 
were given
in Sec.~\ref{LabelSubsectionCircuitModel}.
The difference 
between the gray and yellow gates leads to a staggered pattern in the space-time geometry of the circuits. In this subsection, we focus on this monitored Gaussian circuit with Born-rule measurements. That means we average over all the quantum trajectories of this monitored Gaussian circuit weighted by the Born-rule probability. We numerically simulate this monitored Gaussian circuit using the covariance matrix formulation the technical details of which are reviewed in App.~\ref{app:covariance_matrix}. The Born-rule probability is implemented using an importance sampling scheme for the Kraus operators as explained in App. \ref{LabelSectionAppendixMonteCarloSimulations}.

Upon tuning the parameters $a_{1,2}$ and $b_{1,2}$, this monitored Gaussian circuit exhibits two different phases: (1) the area-law phase and (2) the critical phase. In the area-law phase, the averaged von Neumann EE for a subsystem asymptotes to $O(1)$ values, namely follows the area law, in the long-time limit. In the critical phase, the averaged EE for a subsystem which is an interval is numerically found to saturate to values proportional to the logarithm of the length of the interval in the long time limit. In the remainder of the paper, EE always refers to the von Neumann entanglement entropy. Upon choosing fixed values $a_1=0.5$ and $b_1=1$, we obtained the phase diagram of this monitored Gaussian circuit with Born-rule measurements as a function of $a_2$ and $b_2$ which is displayed in Fig. \ref{fig:Fig-Results-Born-Rule}(a). This phase diagram is obtained by numerically simulating the monitored Gaussian circuit on a Majorana chain with periodic boundary conditions. The boundary between the area-law phase and the critical phase in the phase diagram is identified using the crossing behavior of the two-interval mutual information for different total system sizes $L=64,128,256,512$, as we explain below.

\begin{figure}
    \centering
    \includegraphics[scale=0.65]{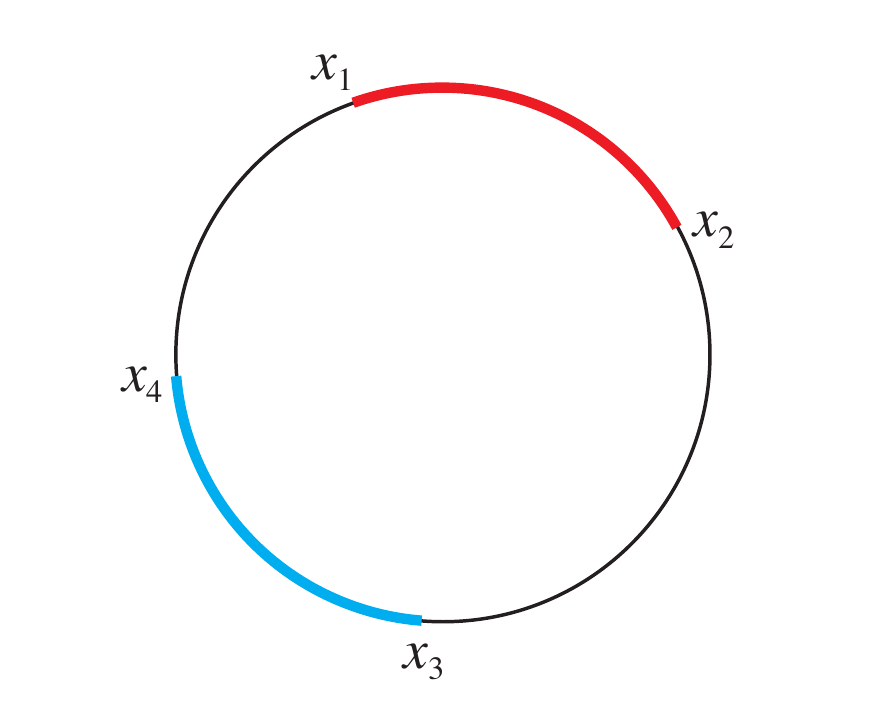}
    \caption{On a length-$L$ Majorana chain with periodic boundary condition, we study the averaged mutual information between the two intervals $[x_1, x_2]$ and $[x_3, x_4]$.}
\label{LabelFigMutualInformation}
\end{figure}

We scan different vertical cuts in the phase diagram, each with a fixed value of $b_2$, to determine the critical value of $a_2$ and extract critical exponents. The phase transition at different values of $b_2$ is expected to share the same universality class, which is consistent with our numerical results. Therefore, in the following, we mostly focus on the vertical cut with $b_2= 1$ (red line in Fig. \ref{fig:Fig-Results-Born-Rule} (a)) as an example. With $b_2=1$ fixed, to accurately determine the value of $a_2$ at the entanglement phase transition, we numerically calculate the averaged mutual information $I$ between the two intervals $[x_1, x_2]$ and $[x_3,x_4]$ (Fig.~\ref{LabelFigMutualInformation}). When $a_2$ is at the transition point, conformal symmetry ensures that the averaged mutual information $I$ only depends on the locations and the lengths of the intervals via the cross-ratio
$\eta = \frac{R(x_2-x_1)R(x_4-x_3)}{R(x_3-x_1)R(x_4-x_2)}$
\cite{Jian2020,Li2020ConformalSym,LiChenFisher2019},
where $R(x) := 
{L\over \pi} \sin\left({\pi\over L} |x|\right)$ is the chord distance. To locate the phase transition point, we choose $x_2 = x_1 + L/8$, $x_3 = x_1 + L/2$, and $x_4 = x_1 + 5L/8$, which yields $\eta =(\sin(\pi/8)/\sin(\pi/2))^2=\sin^2(\pi/8) = \frac{2-\sqrt{2}}{4}$. The averaged two-interval mutual information $I(a_2)$ with  $\eta  =\sin^2(\pi/8) $ is numerically calculated as a function of $a_2$ for various total system sizes $L=64, 128, 256, 512$ and plotted in Fig. \ref{fig:Fig-Results-Born-Rule} (b). The crossing of the averaged mutual information $I(a_2)$ for different $L$ indicates that the phase transition occurs at $a_{2c} = 0.24$. Technically, we average the mutual information $I(a_2)$ over quantum trajectories and all values of $x_1$ (while the spacings between $x_{1,2,3,4}$ are fixed). The averaging over $x_1$ is merely a technical procedure to speed up the convergence. Even if we do not actively average over $x_1$, the mutual information $I$ after averaging over different quantum trajectories only depends on the sizes of and the spacing between the two intervals. Similar averaged-mutual-information calculations at $b_2=1$ for different choices of $\eta$ produce the same value of $a_{2c} = 0.24$.

As is shown in Fig. \ref{fig:Fig-Results-Born-Rule} (c),
finite-size scaling data collapse 
can be
achieved by plotting $|I(a_2)-I(a_{2c})|$ as a function of $(a_2 - a_{2c})L^{1/\nu}$ with an exponent $\nu = 2.1 \pm 0.1$ 
(reported in the summary Table~\ref{Tab:exponent_summary}). To extract $\nu$ and estimate its uncertainty from the scaling data collapse, we apply the algorithm introduced in Ref.~\onlinecite{Skinner2019MIPT}. We give a brief review of this algorithm in App. \ref{app:nu_extraction}.

\begin{figure*}[hbt]
   \centering
 \includegraphics[width=0.95\textwidth]{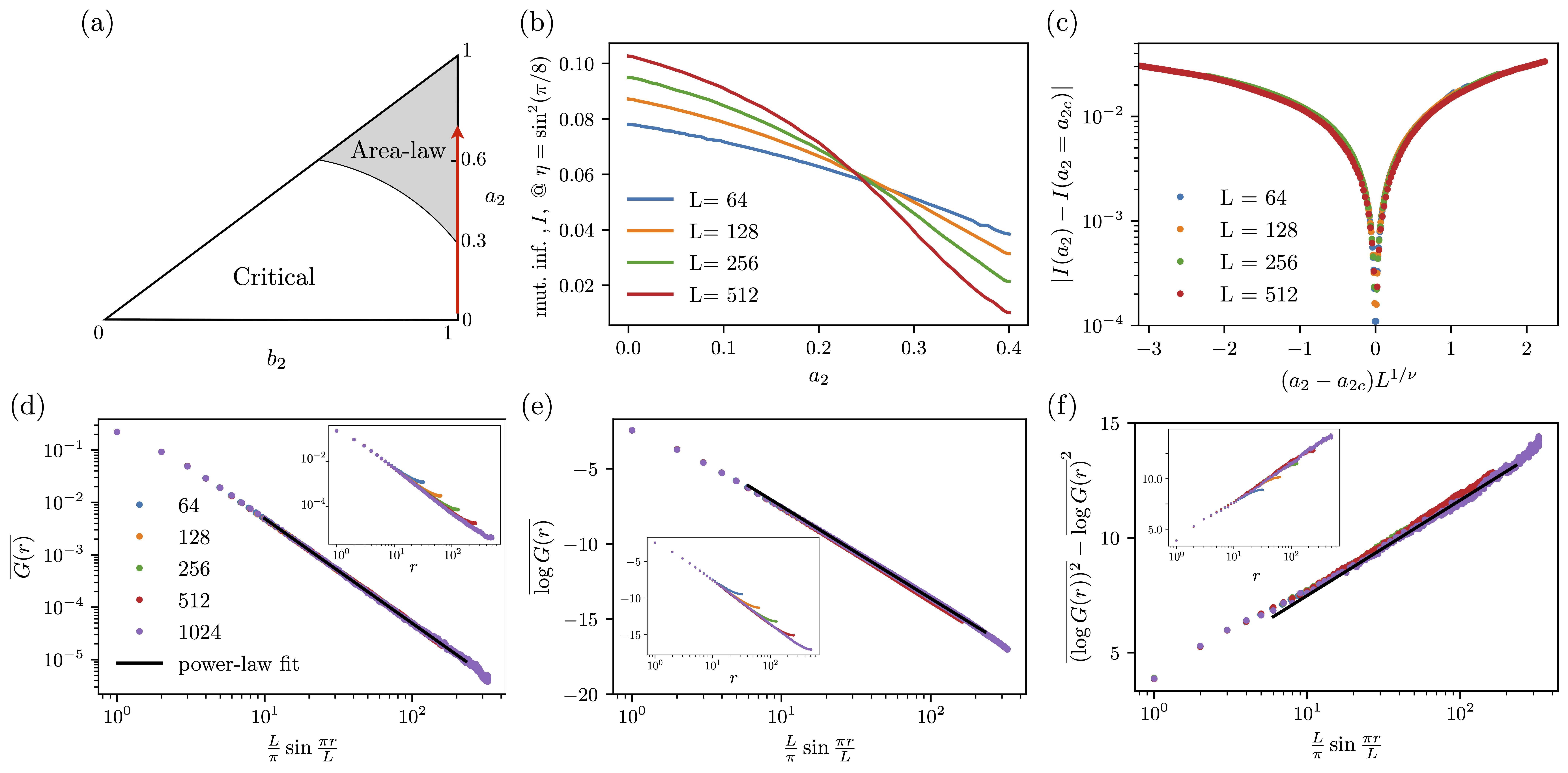}
\caption{We numerically simulate the monitored Gaussian circuit with Born-rule measurements whose Kraus-operator ensemble is given in Sec. \ref{LabelSubsectionCircuitModel}. We fix the parameters $a_1=0.5$ and $b_1=1$ for the numerical study. The phase diagram as a function of $a_2$ and $b_2$ is shown in (a). Upon fixing $b_2=1$, the averaged mutual information $I$ between the two intervals
$[x_1,x_1+L/8]$ and $[x_1+L/2,x_1+5L/8]$ as a function of $a_2$ for different total system sizes $L$ is calculated numerically and presented in (b).
From the crossing of these functions, we identify the value $a_{2c} = 0.24$ where the entanglement phase transition between the critical phase and area-law phase occurs. (c) shows the data collapse when $|I(a_2)- I(a_{2c})|$ is plotted as a function of $(a_2 - a_{2c})L^{1/\nu}$ with $\nu = 2.1\pm 0.1$. (d), (e), and (f) show the dependence of $\overline{G(r)}$, $\overline{\log G(r)}$, and the 2nd cumulant of $\log G(r)$ on 
the chord distance $R(r) = \frac{L}{\pi} \sin \frac{\pi |r|}{L}$.
}
\label{fig:Fig-Results-Born-Rule}
\end{figure*}

Here, we comment on the conformal symmetry of the system. Our numerical results, in particular, the crossing behavior of mutual information shown in Fig. \ref{fig:Fig-Results-Born-Rule} (b), are fully consistent with the conformal symmetry at the transition between the area-law phase and the critical phase.  Away from the transition, i.e. for $a_2 \not = a_{2c}$, the average mutual information is expected to depend not only on the cross-ratio $\eta$, which is fixed to $\eta =\sin^2(\pi/8)$, but also  on the total system size $L$. This is  a reflection of the  lack of conformal symmetry away from the transition. This size dependence is born out in the data of Fig. \ref{fig:Fig-Results-Born-Rule} (b). More specifically, in the area-law phase, where $a_2 > a_{2c}$, the mutual information
decreases with increasing system size $L$, consistent with  crossover to the expected vanishing mutual information in the area-law phase. On the other hand, in the critical phase where $a_2 < a_{2c}$, the mutual information (at fixed
cross-ratio $\eta =\sin^2(\pi/8)$) is seen to increase with system size $L$. The $L$-dependence is the expected behavior in the crossover regime between the transition, and the renormalization group (RG) fixed point characterizing the critical phase. The latter is also known to respect the conformal symmetry \cite{Jian2020}. The increase of the mutual information (at fixed
cross-ratio $\eta=\sin^2(\pi/8)$)
with system size $L$ is indicative of a larger value of the mutual information at the corresponding fixed $\eta$
of the critical phase. (We note that in the analogous cross-over for forced measurements, Fig.~\ref{fig:Fig-Results-No-Born} discussed below, 
the mutual information at the corresponding infrared fixed point is in fact known  from Ref.~\onlinecite{Jian2020}
to be significantly larger than that at the transition, consistent with the increase as a function of $L$.) 
We stress that the independence of the mutual information
of system size $L$ is a property of conformal symmetry, present only at an RG fixed point. In the crossover regime describing the critical phase (with $a_2<a_{2c}$), this independence is only present in the two asymptotic regimes
of system size $L \ll \xi$, and $ L \gg \xi$, where $\xi$ is the crossover length scale set by the distance
$ a_{2c} - a_2 $ from the transition and diverging as $a_{2c} - a_2$ tends to zero. The former limit corresponds to the transition, and the latter to the RG fixed point characterizing the universality
class of the critical phase.

\begin{figure}
    \centering
    \includegraphics[scale=0.8]{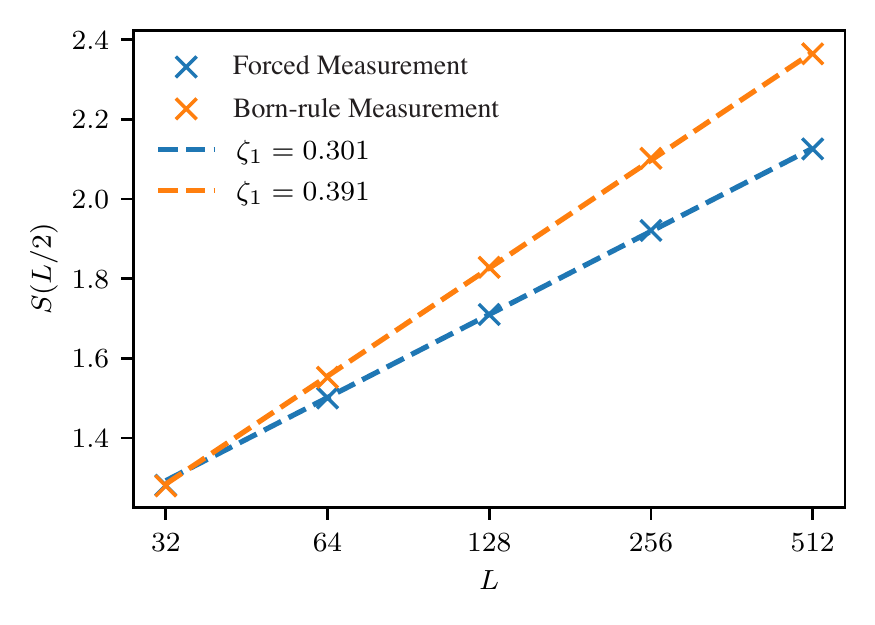}
\caption{Logarithmic scaling of the half-system EE $S(L/2)$ at the entanglement phase transitions presented in Figs.~\ref{fig:Fig-Results-Born-Rule} and \ref{fig:Fig-Results-No-Born}. The total system has a length $L$ and a periodic boundary condition. 
}
    \label{fig:diii-ee-critical}
\end{figure}

At the phase transition ($b_2 = 1$ and $a_2 = 0.24$), following the arguments in Ref. \onlinecite{JianYouVassuerLudwig2020MIPT}, conformal symmetry leads to the logarithmic scaling of the half-system EE $S(L/2) \sim \zeta_1 \log L$ which is confirmed by our numerical results shown in Fig. \ref{fig:diii-ee-critical}. The prefactor $\zeta_1$, which is associated with the universality class of this transition, is found to be  $\zeta_1=0.39\pm 0.02$.

Next, we describe the tool that will allow us to establish very strong evidence
that the entanglement phase transition in the monitored Gaussian circuit with Born-rule measurements and the forced-measurement counterpart belong to different universality classes.
We consider the (squared) Majorana fermion correlation function
\begin{eqnarray}
\label{LabelEqDEFCorrelatorMainText}
G(p, p+r; {\cal C})
=
\left(
\langle {\bf i}
{\hat \gamma}_p {\hat \gamma}_{p+r}\rangle_{\cal C}
\right)^2
\end{eqnarray}
evaluated in the state occurring at the final time slice in the long-time limit.
${\cal C}$ denotes the realization of the Gaussian circuit that corresponds to a quantum trajectory.
Following the correspondence established in Ref. \onlinecite{Jian2020}, the averaging over the quantum trajectories of the Gaussian circuit can be interpreted as the averaging over random disorder in the corresponding static Hamiltonian system. 
The 
{\it logarithm} of such correlation functions in random systems is known to be 
self-averaging~\cite{LudwigHierarchies1990,Pixley2022Multifractality}.
For this reason, we specifically consider the {\it typical} critical exponent characterizing the scaling of
the average of 
$\log G$, as well as the universal fluctuations of $\log G$ about its average. As briefly reviewed in
Appendix \ref{LabelSectionAppendixMultifractal}, at the phase transition
all moments of the correlator in Eq.~\eqref{LabelEqDEFCorrelatorMainText}
turn out to scale with in general
algebraically independent exponents. As a consequence, all {\it cumulants} of the random variable $\log G(p, p+r; {\cal C})$ grow proportional to the logarithm of the chord distance
$R(r) = 
{L\over \pi} \sin\left({\pi\over L} |r|
\right)$, with the coefficients of proportionality all being
(in general different) universal numbers (compare Eqs.~\eqref{LabelEqCumulants} and \eqref{LabelEqCoefficientsOfCumulants}). Since the 
averaged behaviors of $G(p, p+r; {\cal C})$ and the associated cumulants are independent of $p$, we will use $G(r)$ as a shorthand notation of $G(p, p+r; {\cal C})$.

The first {\it cumulant} 
of $\log G$ at the entanglement phase transition should scale as
\begin{align}
\label{LabelEqLogGVerusLogR}
    \overline{[\log G(r)]} \sim  -2 x^{(1)} \log R(r)
\end{align}
with the typical exponent $X_{typ}:= x^{(1)}$.
In this subsection, 
the overbar 
$\overline{~{ {\cdots}^{}}~}$ 
represents the weighted average over all quantum trajectories with respect to the Born-rule probability. Fig. \ref{fig:Fig-Results-Born-Rule} (e) depicts 
this 1st cumulant versus $\log R(r)$, from which we can extract the exponent 
$X_{typ}=$~$x^{(1)}= 2.66 \pm 0.05$. At the phase transition, the second cumulant $\overline{[\log G]^2} - \overline{[\log G]}^2$ should scale as
\begin{align}
\label{LabelEqSecondCumulantVersusLogR}
    \overline{[\log G(r)]^2} - \overline{[\log G(r)]}^2 
    \sim  -2 x^{(2)} & \log R(r)
\end{align}
with the exponent $x^{(2)}$. From Fig. \ref{fig:Fig-Results-Born-Rule} (f), we can extract the exponent $-x^{(2)} = 1.80 \pm 0.04$ 
which describes the universal scaling of
the statistical fluctuations of $\log~G$ about its mean. 
We contrast this with the first moment average $\overline{G}$ 
of $G$ (as opposed to of $\log G$)
which exhibits a power-law decay 
$R(r)^{-2X_1}$
(see Fig. \ref{fig:Fig-Results-Born-Rule} (d))
at the transition with an  exponent that we find to be 
$X_1 = 1.00 \pm 0.02$.
We want to emphasize that at the entanglement phase transition, the exponent $X_1$ for the power-law decay of the {\it average} correlation function $\overline{G}$
is thus found to be significantly different from the exponent $X_{typ}$ describing the decay of the {\it typical} correlation function, $X_1 = 1.00 \pm 0.02 < X_{typ}= 2.66 \pm 0.05$. This is a reflection of the rich  scaling behavior of the correlation function that is usually referred to as {\it multifractal} (\cite{Pixley2022Multifractality,LudwigHierarchies1990},
and Appendix \ref{LabelSectionAppendixMultifractal}).
All the exponents are summarized in Table \ref{Tab:exponent_summary}.

We conclude this section by commenting on the critical phase.
First, as
already stated in the first paragraph of Sec.~\ref{LabelSubsectionBornRuleMeasurements},
the averaged half-system EE $S(L/2)$ in the critical phase
is numerically found to be  proportional to the logarithm of subsystem size, a reflection of criticality (in the sense that the phase is governed by an RG fixed point with conformal symmetry).
Second, we note that in the critical phase the 
average
and typical values of the correlation functions $G(r)$ are expected to scale with the  {\it same} power
$\propto R(r)^{-2 X}$ with $X=1$ (up to logarithmic corrections to scaling).
This will follow from the discussion in Sec.~\ref{LabelSubSectionFieldTheory}.
This ``self-averaging'' property of $G(r)$  in the critical phase is in sharp contrast
with the rich scaling properties of the same correlation function at the transition, where the average
and typical correlation functions scale with vastly different exponents (as discussed in the paragraph above, and summarized in Table~\ref{Tab:exponent_summary}). Analogous rich scaling behavior of the correlation function at the entanglement transition also occurs for the monitored Gaussian circuits with forced measurements, as discussed in the following section. As pointed out in Ref.~\onlinecite{Jian2020}, $G(r)$ in the critical phase with forced measurement is also self-averaging, like the case with Born-rule measurements.

\subsection{Forced measurements}
\label{LabelSubsectionForcedMeasurements}

Next, we numerically simulate the monitored Gaussian circuit (with the geometry as in
Fig.~\ref{fig:diii-circuit} (a)) using the {\it forced-measurement} 
statistical weights for each
quantum trajectory discussed at the end of
Sec. \ref{LabelSubsectionCircuitModel}, and Eq.~\eqref{LabelEqForcedMeasurementWeight}.
Similar to the case with Born-rule measurements, we simulate this monitored Gaussian circuit using the covariance matrix formulation of which the technical details are reviewed in App.\ref{app:covariance_matrix}. For each gate in the circuit, the measurement outcome $\Vec{n}_{A/B}$ (parameterized by $s$ and $\varphi$) is randomly sampled following the statistical weight specified in Eq. \eqref{LabelEqForcedMeasurementWeight}.

\begin{figure*}[htb]
\centering
 \includegraphics[width=0.95\textwidth]{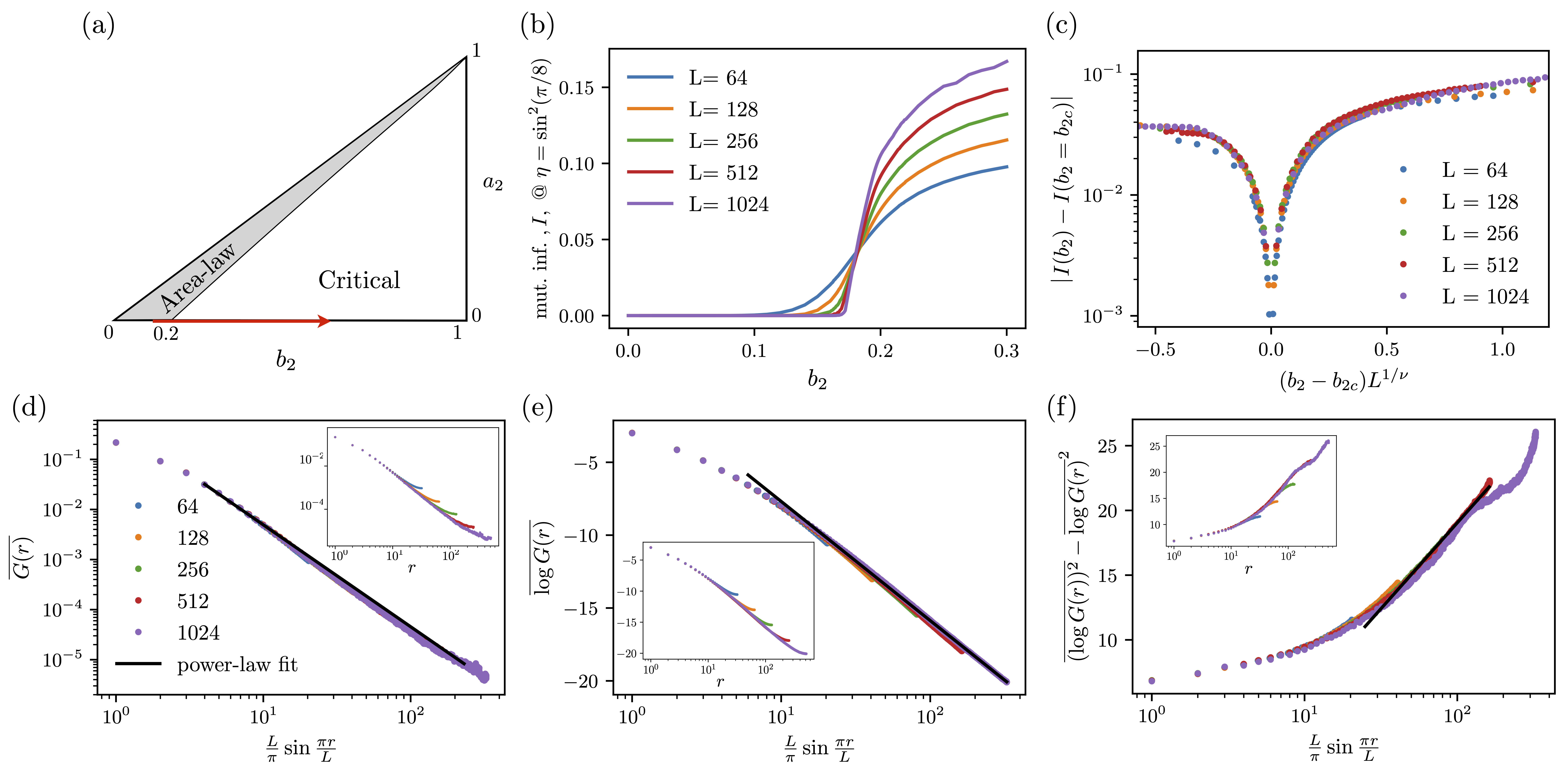}
\caption{
We numerically simulate the monitored Gaussian circuit with forced measurements whose Kraus-operator ensemble is given in Sec. \ref{LabelSubsectionCircuitModel}. We again fix the parameters $a_1=0.5$ and $b_1=1$ for the numerical study. The phase diagram as a function of $a_2$ and $b_2$ is shown in (a). 
Upon fixing $a_2=0$, 
the averaged mutual information $I$ between the two intervals
$[x_1,x_1+L/8]$ and $[x_1+L/2,x_1+5L/8]$ as a function of $b_2$ for different total system sizes $L$ is calculated numerically and presented in (b). From the crossing point of these functions, we identify the value $b_{2c} = 0.18$ where the phase transition between the critical phase and area-law phase occurs. (c) shows the the data collapse when $|I(b_2)- I(b_{2c})|$ is plotted as a function of $(b_2 - b_{2c})L^{1/\nu}$ with $\nu = 1.9\pm 0.1$. (d), (e), and (f) show the dependence of $\overline{G(r)}$, $\overline{\log G(r)}$, and the 2nd cumulant of $\log G(r)$ on 
the chord distance $R(r) = \frac{L}{\pi} \sin \frac{\pi |r|}{L}$.}
    \label{fig:Fig-Results-No-Born}
\end{figure*}

Upon tuning the parameters $a_{1,2}$ and $b_{1,2}$ which, as before,  changes the staggered pattern in the space-time of the circuit between gray and yellow gates in Fig.~\ref{fig:diii-circuit} (a), the circuit
again exhibits an area-law and a critical phase. Choosing fixed values $b_1=1$ and $a_1=0.5$ (which is the same as in Sec. \ref{LabelSubsectionBornRuleMeasurements}), 
the phase diagram as a function of $a_2$ and $b_2$ is displayed in Fig.~\ref{fig:Fig-Results-No-Born} (a). Similar to the Born-rule case, we use the crossing of the averaged mutual information to determine the boundary between the 
area-law phase and the critical phase. Notice that switching from the Born-rule measurement to the forced measurement leads to a significant change in the phase boundary. More importantly, we show below that the universality class of the forced-measurement-induced entanglement phase transition differs from that of the Born-rule counterpart.

We scan different horizontal cuts in the phase diagram, each with a fixed value of $a_2$, to determine the critical value of $b_2$ and extra critical exponents. The phase transition for different values of $a_2$ belongs to the same universality class. In the following, we mostly focus on the horizontal cut with $a_2 = 0$ (red line in Fig. \ref{fig:Fig-Results-No-Born} (a)) to investigate the critical behavior at the phase transition. As before, we numerically compute the averaged mutual information $I$ between two intervals $[x_1,x_2=x_1+L/8]$ and $[x_3=x_1+L/2, x_4= x_1+5L/8]$ 
to accurately determine, this time, the value of $b_2$ at the phase transition. Again, the corresponding cross-ratio is fixed at $\eta= \sin^2(\pi/8)$. The averaged mutual information $I(b_2)$ is numerically calculated as a function
of $b_2$ for various total system sizes $L=64, 128, 256, 512, 1024$ and plotted in
Fig.~\ref {fig:Fig-Results-No-Born} (b). Using the crossing of the mutual information for different system sizes $L$, we conclude that the entanglement phase transition occurs at $b_{2,c}=0.18$.

As shown in Fig.~\ref {fig:Fig-Results-No-Born} (c), finite-size scaling data collapse  can be achieved by plotting $|I(b_2)-I(b_{2c})|$ as a function of $(b_2-b_{2c}) L^{1/\nu}$ with an exponent $\nu=1.9\pm 0.1$ (reported in the summary Table~\ref{Tab:exponent_summary}).
The method to extract $\nu$ and estimate its uncertainty is the same as in the case with Born-rule measurements.

It is worth noting that upon
applying the correspondence established Ref. \onlinecite{Jian2020}, this forced-measurement-induced entanglement transition under discussion here is equivalent to the thermal-metal-insulator transition
in a two-dimensional disordered non-interacting fermionic Hamiltonian system in symmetry class 
DIII~\footnote{Symmetry class 
DIII describes the Bogoliubov-deGennes Hamiltonian of a time-reversal invariant superconductor (or insulator with charge conservation).}. 
The latter has been numerically studied using a Chalker-Coddington-model-based approach in Ref. \onlinecite{Fulga2012DIIIMIT}. 
In this study, the
critical exponent $\nu$ for this thermal metal-insulator transition was found to be $\nu \approx 2$ which is consistent with our result. Ref. \onlinecite{Jian2020} also studied this transition in the context of Gaussian circuits but did not include a detailed analysis of the exponent $\nu$ and the other 
exponents we introduce in the rest of this section.

At the phase transition ($a_2 = 0$ and $b_2 = 0.18$), conformal symmetry leads to the logarithmic scaling of the half-system EE $S(L/2) \sim \zeta_1 \log L$, which is again confirmed by our numerical results shown in Fig. \ref{fig:diii-ee-critical}. The prefactor $\zeta_1$, which is associated with the universality class of this transition, is found to be  $\zeta_1=0.30\pm 0.04$. 

Next, we consider again the correlation function
$G$ defined in
Eq.~\eqref{LabelEqDEFCorrelatorMainText}. We stress again that, in this subsection, we calculate the averaged behavior $\overline{G}$, $\overline{\log G}$
and so on using the forced-measurement statistical weight, which assigns equal weights to all quantum trajectories as discussed at the end of Sec. \ref{LabelSubsectionCircuitModel}. At the phase transition, the first {\it cumulant} of $\log G$ scales as in Eq.~\eqref{LabelEqLogGVerusLogR}, which agrees with our numerical result plotted in Fig.~\ref{fig:Fig-Results-No-Born} (e). The extracted typical
exponent is $X_{typ}=$~$x^{(1)}=3.53 \pm 0.04$. The second {\it cumulant}
of $\log G$ should scale as in Eq.~\eqref{LabelEqSecondCumulantVersusLogR} at the transition. From Fig.~\ref{fig:Fig-Results-No-Born} (f)
we extract the value of the exponent $-x^{(2)}= 5.7 \pm 0.2$,
describing the universal scaling of
the statistical fluctuations of $\log G$ about its mean. We contrast this again with the first moment 
average $\overline{G}$ of $G$ (as opposed to of $\log G$)
which exhibits a power-law decay 
$R(r)^{-2X_1}$
(see Fig. \ref{fig:Fig-Results-No-Born} (d))
at the transition with an exponent that we find to be 
$X_1 = 1.02 \pm 0.03$. We want to emphasize again that, at the entanglement transition, the exponent $X_1$ for the power-law decay of the {\it average} correlation function $\overline{G}$
is found to be significantly different from the exponent $X_{typ}$ describing the decay of the {\it typical} correlation function, $X_1 = 1.02\pm 0.03 < X_{typ}= 3.53 \pm 0.04$.
This is analogous to the case of Born-rule measurements (while the value of $X_{typ}$ is very different), and is again a reflection of the rich  scaling behavior of the correlation function that is usually referred to as {\it multifractal} 
(see Refs.~\onlinecite{LudwigHierarchies1990,Pixley2022Multifractality}
and App. \ref{LabelSectionAppendixMultifractal}).

We conclude this section by commenting on the critical phase. 
First, as established numerically in Ref.~\onlinecite{Jian2020},
the averaged EE $S(L/2)$ in the critical phase has a logarithmic scaling as a function of $L$, i.e. $S(L/2) \sim \log L$, reflecting the conformal symmetry of the critical phase.
Second, in that same reference,  
the 
average
and typical values of the correlation functions  $G(r)$ 
were found to scale with the  {\it same} power
$\propto R(r)^{-2 X}$ with $X=1$ (up to logarithmic corrections to scaling).
This also follows  from the theoretical understanding of the critical phase discussed in Ref.~\onlinecite{Jian2020} (see also
Sec.~\ref{LabelSubSectionFieldTheory}).
This ``self-averaging'' property of $G(r)$  in the critical phase is in sharp contrast
with the rich scaling properties of the same correlation  function at the entanglement transition, where the average and typical correlation functions scale with vastly different exponents (as discussed in the paragraph above, and summarized in Table~\ref{Tab:exponent_summary}).
As discussed in the preceding section,
analogous different scaling behavior of the correlation function at the transition and in the critical phase also occurs for monitored Gaussian circuits with Born-rule measurements.

All the exponents are summarized in Table \ref{Tab:exponent_summary}.

\begin{table*}[]
    \centering 
\renewcommand{\arraystretch}{1.2}
\begin{tabular}{c|ccc}
    \hline
    Exponents & $\ \ $ Born-rule measurement $\ \ $ & $\ \ $Forced  measurement$\ \ $ & $\ \ $Quantities used to extract the exponents$\ \ $ \\ 
    \hline
    \hline
    $\nu$ & $2.1\pm 0.1$  & $1.9\pm 0.1$ & Mutual information, $I$ \\
    $X_1$ & $1.00\pm 0.02$  & $1.02 \pm 0.03$ &  Majorana fermion correlation function, $\overline{G(r)}$ \\
    $x^{(1)}$ & $2.66 \pm 0.05$  & $3.53 \pm 0.04$  & $\overline{\log G}$  \\
    $-  x^{(2)}$ & $1.80\pm 0.04$  & $5.7\pm 0.2$  & Second cumulant of ${\log G}$  \\
    $\zeta_1$ & $0.39\pm 0.02$  & $0.30 \pm 0.04$ & Coefficient of entanglement entropy, $S(L/2)$  \\
    \hline
\end{tabular}
\caption{This table summarizes the critical exponents at the phase transition between the critical phase and the area-law phase of the monitored Gaussian circuit with Born-rule and forced measurements.
Here, $X_{typ}:= x^{(1)}$ is the typical exponent.}
\label{Tab:exponent_summary}
\end{table*}

\subsection{Comparison between Born-rule and forced measurements}

\label{LabelSubsectionComparisonBornForcedMeasurements}

Amongst the numerical results presented in the two preceding subsections
Sec. \ref{LabelSubsectionBornRuleMeasurements} and Sec. \ref{LabelSubsectionForcedMeasurements} and summarized in Table \ref{Tab:exponent_summary}, the typical exponent $X_{typ}=x^{(1)}$ and the exponent $x^{(2)}$,
which are
associated with the 1st and the 2nd cumulants of $\log G$, are widely different (beyond error bars) between the phase transitions in the Gaussian circuits with Born-rule and forced measurements. Such 
a difference offers strong evidence that the phase transitions with Born-rule and with forced measurements belong to different universality classes.

The coefficient $\zeta_1$ of the half-system EE as a function of $\log L$ is also different between the Born-rule-measurement-induced transition and forced-measurement-induced transition. For Born-rule measurements, the value $\zeta_1=0.39 \pm 0.02$, whereas for forced measurement $\zeta_1=0.30 \pm 0.04$ is found. We note that the correlation length exponent $\nu$ appears to differ only very little between the two transitions.

Besides the very strong numerical evidence for the difference between these two universality classes, there are theoretical arguments that go hand in hand with these numerical observations.  
Owing to the correspondence established in Ref.~\onlinecite{Jian2020}, we know the field theory (or the ``statistical mechanics model") underlying both transitions, which follows, as already mentioned above, solely from the AZ symmetry class associated with the circuit. This will be discussed in the following section.

\section{Theoretical Description}
\label{sec:replica_theory}

\subsection{Field theory and statistical mechanics model}

\label{LabelSubSectionFieldTheory}

In this section, we provide a theoretical understanding 
of the monitored Gaussian circuits for which the numerical results were  presented in the previous section, in terms of the underlying  field theories (or equivalently the ``statistical mechanics models'') describing these two entanglement transitions in monitored Gaussian circuits with Born-rule and forced measurements, as well as their adjacent phases. We stress again that the corresponding field theory/statistical-mechanics model is directly determined by the corresponding AZ symmetry class of the monitored Gaussian circuit. As mentioned, a generic monitored Gaussian circuit preserving only fermion parity belongs to symmetry class DIII.

{\it Forced measurement} -  It is simplest to start with the forced measurement case. 
Let us first recall that, in general, a circuit with forced measurements is equivalent to a random tensor network (RTN) \cite{Jian2020,Nahum2021AlltoAll}. In the language 
of the corresponding disordered static  Hamiltonian system whose spatial coordinates describe
the space-time of the circuit, the staggered pattern of the Gaussian circuit, associated with the difference 
between the $A$ and the $B$ sublattices, 
corresponding to the yellow and the gray gates
in Fig. \ref{fig:diii-circuit} (a),
will favor the topologically trivial 
or the topological non-trivial superconductor, both thermal insulators in symmetry class DIII in two spatial dimensions, depending on the 
``direction'' of 
staggering (see   Ref.~\onlinecite{Fulga2012DIIIMIT}).
Both, the topologically trivial and the topologically non-trivial phases
correspond to an area-law phase of the monitored Gaussian circuit. 
There is an intervening critical phase that appears for intermediate
staggering~\cite{Fulga2012DIIIMIT,Jian2020}. This critical phase 
of  the circuit acting on the $d=1$ dimensional Majorana chain is known to be described~\cite{Jian2020,SchnyderRyuFurusakiLudwigClassificationPhysRevB2008,RyuSchnyderFurusakiLudwigNJPhys2010} 
by a so-called principal chiral non-linear sigma model (NLSM) in $(d+1)=2$-dimensional space 
whose field variable
$O({\vec r})\in \SO(n)$ is a special $n\times n$ orthogonal 
matrix,  
in the limit where $n \to 0$~\footnote{ The mechanism for the appearance of the critical phase
in symmetry class DIII, as explained in Ref.~\onlinecite{Fulga2012DIIIMIT}, 
following  Ref.~\onlinecite{ChalkerEtAlThermalMetalPRB65} for class D, 
results in the fact that in this phase the field variable $O({\vec r})$
resides in the component of the orthogonal group $\OO(n)$ which is connected to the identity, i.e. in $\SO(n)$. Therefore, the critical phase is described by the $\SO(n)$ and not by the $\OO(n)$ principal chiral NLSM. As discussed below, the transition out of this critical phase into the adjacent area-law phases is driven by proliferation of topological defects in the $\SO(n)$ principal chiral NLSM}.
This NLSM is formulated using the replica trick where $n$ is the number of replicas of the monitored Gaussian circuit, and we refer to it
as the principal chiral  $\SO(n)$ NLSM.  This is the statistical mechanics model for the circuit.
The Boltzmann weight of this 
NLSM
is $\exp{\{ -S \}}$
which, when written in continuum notation, reads
\begin{eqnarray}
\label{LabelEqPCMAction}
S= \int d^2 {\vec r} \ {1\over 2g} \Tr \left [   ({\bf \nabla}O^{-1}) ({\bf \nabla} O)\right  ].
\end{eqnarray}
This Boltzmann weight manifestly exhibits
$\SO(n) \times \SO(n)$
continuous symmetry corresponding to right- and left-multiplication
with spatially independent group elements in $\SO(n)$. 
The critical phase of the circuit corresponds to the phase of the NLSM where the $\SO(n) \times \SO(n)$
symmetry is spontaneously broken to the 
diagonal
$\SO(n)$
symmetry where right and left multiplications are no longer independent but implemented,
respectively, by the same 
matrix and its inverse. 
This phase is described by $n(n-1)/2$
(non-interacting)  Goldstone modes of this spontaneous symmetry breaking in the replica limit $n\to 0$~\footnote{Owning to the $n \to 0$ limit (or a corresponding supersymmetric formulation, briefly alluded to below), the Mermin-Wagner theorem is known not to apply.}. In the context of the corresponding disordered Hamiltonian system of non-interacting fermions in one spatial dimension higher, the critical phase of the Gaussian circuit/Gaussian RTN corresponds to a disordered thermal metal in two spatial dimensions.
The transition out of the critical phase into each one of these two
area-law phases is driven by proliferation
of topological defects in the NLSM field variable $O({\vec r})\in \SO(n)$.
Note that both transitions are known to be in the same (bulk) universality class in two spatial dimensions \cite{Fulga2012DIIIMIT}. One of these transitions was studied explicitly in the Gaussian circuit in Ref.~\onlinecite{Jian2020}.
The type of topological defects is dictated by the fundamental group of the target space 
$\SO(n)$, which is known to be $\pi_1(\SO(n)) = \mathbb{Z}_2$ for generic $n$ (unless $n=2$ where it is $\mathbb{Z}$).
The presence of topological defects is represented by an additional term in Eq. \eqref{LabelEqPCMAction}
(not written explicitly), whose coupling constant is the defect fugacity.
The specific theoretical description of this transition
for the current NLSM  (which, as mentioned, is in symmetry class DIII)  proceeds in very close analogy to the discussion of the metal-insulator transition in the ``symplectic" 
(or ``spin-orbit'') symmetry class AII in Ref. \onlinecite{FuKane2012Sympletic} in the same spatial dimension, while the specific target space of the latter NLSM is different.
The technical details of the generalization of this analysis to the current DIII symmetry class will be presented in separate follow-up work.
Here, in order to be slightly more specific, we note that the system discussed in the present paper may be viewed in a sense as a superconductor version of the one discussed in Ref. \onlinecite{FuKane2012Sympletic} (which is in a different AZ symmetry class). The phase diagrams of both systems feature a parameter that tunes between a  $\mathbb{Z}_2$ topologically non-trivial and a topologically trivial phase with an intervening critical (metallic) 
phase~\footnote{ E.g., compare for the system discussed in the present paper,
Fig. 3 of Ref. \onlinecite{Fulga2012DIIIMIT}, 
with the phase diagram in Fig. 11 of the symplectic class in Ref. 
\onlinecite{ObuseFurusakiETAL2007}
(with tuning parameter $x$).}.
In both systems,  
one can tune through the transition out of the critical (metallic) phase either with this tuning parameter or by varying the strength
of disorder.

{\it Born-rule measurement} - 
The entanglement transition with Born-rule measurements, on the other hand, is described
by the $n\to 1$ limit of the same principal chiral $\SO(n)$ NLSM
(as opposed to the $n\to 0$ limit 
in the case of forced measurements). 
In the following we present two arguments (i) and (ii) that lead to the conclusion that the 
principal chiral $\SO(n)$ NLSM in the replica limit $n\rightarrow 1$ describes the entanglement transition with Born-rule measurements. Both arguments are essential for reaching this conclusion.
(i) First, as established in Ref. \onlinecite{JianYouVassuerLudwig2020MIPT} (see also Ref. \onlinecite{BaoChoiAltman2020}),
for Born-rule measurements the replica limit $n \to 1$ has to be taken.
The ``extra" replica (as compared to the forced measurement case)  accounts for the Born-rule probability 
providing the statistical weight for each quantum trajectory.
(ii) Second, it is important that, in the $n\to 1$ limit, the partition function of the statistical-mechanics model/field theory 
goes to a constant (independent of the system size). Following the argument in Ref. \onlinecite{JianYouVassuerLudwig2020MIPT}, the partition function of the statistical mechanics model in this limit is 
directly related to the total Born-rule probability of all the quantum trajectories. This total Born-rule probability is guaranteed to be unity
due to the POVM condition satisfied by the Kraus operator ensemble of each measurement regardless of the total system size. (We refer the readers to Sec. \ref{LabelSectionMonitoredGaussianCircuitsGeneral} and Eq.~\eqref{eq:POVM_nvec} for details on the POVM conditions for the monitored Gaussian circuit.)
The condition that the partition function goes to a constant in the replica limit $n\rightarrow 1$ is satisfied by the principal chiral $\SO(n)$
NLSM because the number
$n(n-1)/2$ of degrees of freedom (describing the Lie algebra of $\SO(n)$) of this NLSM, which is equal to the dimension of the special orthogonal group manifold $\SO(n)$,
goes to zero in the $n\to 1$ limit. This implies, as required, a constant partition function (and a vanishing conformal central charge at the transition) in this replica limit.

Moreover, the RG beta function for the coupling constant $g$ of this NLSM is 
known
to be proportional to $(n-2)$ at weak coupling, namely at
small $g$
(see e.g. Refs.~\onlinecite{PolyakovBook1987} or \onlinecite{Jian2020}).
This result implies the stability of the weak coupling fixed point of the NLSM, described by the $n(n-1)/2$ Goldstone modes of the spontaneous symmetry breaking, in both replica limits $n\rightarrow 0 $ and $n\rightarrow 1$. The stable weak coupling fixed point in the limit
$n\rightarrow 0 $ represents the critical phase of the monitored Gaussian circuit with forced measurements discussed above and in Ref.~\onlinecite{Jian2020}. The stable weak coupling fixed point 
in the limit $n\rightarrow 1 $ represents the critical phase of the monitored Gaussian circuit with Born-rule measurements. \footnote{The coupling constant $g$ of the NLSM is marginally irrelevant in both limits $n\rightarrow 0$ and $n \rightarrow 1$.}

In complete analogy to the case of the $n\to 0$ limit discussed above, this NLSM also has a transition out of its stable critical phase in the $n\to 1$ limit, which is driven again by $\mathbb{Z}_2$ topological defects.
This transition describes the entanglement transition with Born-rule measurements.
The different replica limit $n\rightarrow 1$ for the Born-rule measurement case explains that our numerical results presented in Sec.~\ref{LabelSectionNumericalResults}
observed a different universality class as compared to the one obtained in the forced measurement case in the replica limit $n \rightarrow 0$. Both of the entanglement transitions with Born-rule and forced measurements can be accessed by a perturbative RG treatment controlled (at least to low order) by a small parameter $0  < \epsilon=2-n$, 
which parallels the RG treatment
in Ref. \onlinecite{FuKane2012Sympletic}
of the metal-insulator transition in the two-dimensional symplectic symmetry class AII. 
As mentioned above,
the technical details of this generalization of the analysis
given in Ref. \onlinecite{FuKane2012Sympletic} to the current DIII symmetry class will be presented in separate follow-up work.

We stress again that the field theory/statistical mechanics model identified above
(the principal chiral NLSM on the $\SO(n)$ group manifold) describes both universality classes of 
entanglement transitions, with forced measurements and with Born-rule measurements, in 
the different replica limits.
This field theory is dictated solely by the AZ symmetry class (here DIII). \footnote{We note in passing that, as is well known, the replica limits $n\to 0$ and $n\to 1$ can be avoided by using a different formulation in terms of supersymmetry[Ref(Efetov)].}.

\subsection{Comparison with the loop model with crossings}
\label{LabelSubsectionComparisonLoopModelWithCrossings}

There is a very special and highly fine-tuned version of the monitored Gaussian circuit of Majorana fermions discussed 
in the present paper. 
In this fine-tuned version, every 2-site gate  is drawn (up to unimportant global multiplicative factors), using the Kraus operator language of Eqs.~(\ref{LabelEqKvecn},\ref{LabelEqProbps}),
from the ensemble that contains only (i): the identity operator ($\Vec{n} = (0,1,0)$), (ii): the ``swap gates" ($\Vec{n} = (0,0,\pm 1)$), and (iii): the projective measurement of the local fermion parity $\i \hgamma_i \hgamma_{i+1}$ ($\Vec{n} \rightarrow ( \pm 1,0,0)$).
This monitored Gaussian circuit
 was shown to be described by a two-dimensional statistical mechanics model called the loop model with crossings \cite{NahumSkinnerMajDefect2020FF}. Hence, we refer to this monitored Gaussian circuit
 in the following
as the loop-model-based 
circuit~\footnote{There is also a simpler loop model 'without crossings', where the ``swap gates'' (ii) are absent. This model has no critical phase, but instead an isolated  transition between two area law phases which is in the well-studied universality class of two-dimensional
percolation. Its universality class is manifestly different from the transitions discussed in the present paper, and we will not come back to this simpler loop model in this paper.}.
At the level of entanglement, the loop-model-based circuit (with Born-rule measurements) behaves like a random tensor network, which is {\it not} a general property of a more generic monitored Gaussian circuit (as shown in the previous sections of this paper). The loop-model-based circuit also exhibits 
a measurement-induced critical-to-area-law entanglement phase transition.  In the following, we provide two types of arguments, one numerical and one analytical,
showing that this entanglement phase transition in the loop-model-based circuit 
(which, as mentioned above, is a fine-tuned version of the monitored Gaussian circuit)
is in a universality class different from both entanglement transitions with forced and Born-rule measurements  in the generic Gaussian circuit discussed in the preceding sections of the present paper.

For the numerical arguments, we compare  critical exponents. Numerical values for the correlation length exponent in the loop model were reported in 
Ref. \onlinecite{Nahum2013LoopWithCrossing}
to be $\nu=2.745(19)$ and $\nu=2.87(10)$ at two different points on the line of phases transitions
(governed by the same universality class), which correspond to the critical-to-area-law entanglement phase transitions in the loop-model-based circuit.
In the monitored Gaussian circuit we study in the present paper, we found, as reported
in Sec.~\ref{LabelSectionNumericalResults} and Table~\ref{Tab:exponent_summary}, values for the correlation length exponents 
$\nu=2.1\pm 0.1$ for the Born-rule-measurement-induced transition and $\nu= 1.9\pm 0.1$ for the forced-measurement-induced transition.
Both values are significantly different from the values for $\nu$ for the loop model with crossings listed above. 

We can also compare the universal coefficient $\zeta_1$ of the logarithm of the subsystem size of the EE. 
For the circuits discussed in the present paper, we found
(again as reported
in Sec~\ref{LabelSectionNumericalResults} and Table~\ref{Tab:exponent_summary}) the values
$\zeta_1=0.39 \pm 0.02$ (measurements satisfying Born rule) and $\zeta_1=0.30\pm 0.04$ (forced measurements), which are different (within error bars) from the corresponding values reported in 
Refs.~\onlinecite{Sang2021Negativity,Nahum2013LoopWithCrossing}
 to be $\zeta_1= 2*0.225=
0.45$.

We now come to the analytical arguments.
There is a manifest difference in symmetry: The principal chiral $\SO(n)$ NLSM
field theories with weight Eq.~\eqref{LabelEqPCMAction}
discussed in the present paper (Sect.~\ref{LabelSubSectionFieldTheory}) possess a global $\SO(n) \times \SO(n)$ symmetry
(in the limits $n\to 0$ and $n\to 1$, for forced and Born-rule measurements, respectively), whereas the loop model with crossings is known 
to have only global $\SO(n)$
symmetry and requires taking the replica limit $n\to 1$ \cite{Nahum2013LoopWithCrossing,NahumChalkerVortexLinesPRE85}.
Moreover,  intimately related to these different symmetries, the target space of the former models is a group manifold (namely $\SO(n)$)
whereas the target space of the latter model is a coset space  \cite{Nahum2013LoopWithCrossing,NahumChalkerVortexLinesPRE85}. This coset space for the loop model is the real projective space ${\rm RP}^{n-1}=$ $S^{n-1}/\mathbb{Z}_2$ in the limit $n\to 1$, where $S^{n-1} ={\rm SO}(n)/{\rm SO}(n-1)$ is the unit sphere in $n$-dimensional Euclidean space, which is not a group.

For the entanglement transitions discussed in the present paper 
as well as that in the loop-model-based circuits, one can alternatively choose a supersymmetric (SUSY) formulation~\cite{BookEfetov_1996}
in which the replica limit is avoided. 
All the above statements have an exact counterpart in the SUSY formulation.
In particular, the field 
theories for the entanglement transitions discussed in the present paper, for which details of the SUSY formulation will be presented in separate follow-up work, both possess 
${\cal G}\times {\cal G}$ symmetry, where ${\cal G}$ is a so-called Lie-supergroup (different for
Born-rule and forced measurement cases). 
The loop model with crossings, on the other hand, is known~\cite{Nahum2013LoopWithCrossing,NahumChalkerVortexLinesPRE85}
in the SUSY formulation to be invariant under the lower (super-) symmetry ${\cal G}$ (as opposed to 
${\cal G} \times {\cal G}$). Given that symmetries of the underlying field theories are an identifying characteristic of universality classes, 
one does not expect
any simple
relationship between universality classes
of entanglement transitions
in the loop-model-based circuits and the entanglement transitions of the more generic monitored Gaussian circuits discussed in the present paper.
Technical details of the SUSY formulation of  the theories under consideration here will be presented in separate follow-up work.

We close this section by remarking on another difference which has a simple reflection in physical observables when we compare the loop-model-based circuits and the more generic monitored Gaussian circuit in the preceding sections of this paper. As stressed in Ref.~\onlinecite{Jian2020}, in a loop-model-based circuit, the square of any Majorana-fermion two-point function $G$ (as defined in 
Eq.~\eqref{LabelEqDEFCorrelatorMainText}) can take only values zero or one in a fixed realization of the Gaussian circuit. This implies, in particular, that the $N$-th moment average $\overline{G^N}$ of such a correlation function over quantum trajectories is equal to the first moment $N=1$.
This behavior of the loop-model-based circuits is in sharp contrast with the rich scaling behavior of the correlation function $G$ in the more generic monitor Gaussian circuits discussed in Sec.~\ref{LabelSectionNumericalResults} of this paper (both, for Born-rule and forced measurements). Recall (from Table~\ref{Tab:exponent_summary}) that the
critical exponents describing the decay of the typical ($X_{typ}$) and the averaged (1st moment, 
$X_1$) correlation function
are vastly different.
More generally, all moments $\overline{G^N}$ 
will scale with independent critical exponents
~\footnote{The latter property
represents, as briefly summarized in Appendix~\ref{LabelSectionAppendixMultifractal},
a wealth of  universal data. These  can also be encoded in a scaling form of the probability distribution characterized by a universal function~\cite{LudwigHierarchies1990,Pixley2022Multifractality}.}. 
As already mentioned, such scaling behavior is referred to as {\it multifractal}, and it implies
a continuum of correlation function exponents 
(which appear when continuous values of moment order $N$ are taken).
It turns out that this
rich scaling behavior appearing at
both entanglement phase transitions discussed in the present paper
is a consequence of the non-compactness of the target space (in the SUSY formulation) of the NLSMs for the monitored Gaussian circuits discussed in the present paper. On the other hand, the absence of such rich scaling behavior 
at the entanglement transition of the circuits described by the loop model with crossings and in the corresponding statistical mechanics model is a consequence of the compactness of the target space (in the SUSY formulation) of the corresponding 
NLSMs.
~\footnote{The sole fact that one model contains only critical exponents describing the average (or a few low-order moments) of an observable while another model describes rich scaling behavior featuring unequal average and typical exponents of the same observable, does not in itself imply that the transitions in the two models are in unrelated universality classes.  However, while two such models could lie in the same universality class if the former model forms a subsector of the latter model containing just a subset of the critical exponents of the latter model (as it is the case in the model discussed in Ref.~\onlinecite{GruzbergLudwigReadPRL1999}),
such a situation is {\it not} realized in the case discussed in 
the present paper because the NLSM for the loop-model with crossings is invariant under a different symmetry than
the NLSM for the generic monitored Gaussian circuit discussed in this paper, as detailed above in this Section.
}

\section{Conclusions and Outlook}
\label{sec:conclusions}

In this paper, we studied the entanglement transition in generic  monitored Gaussian circuits 
with no symmetry other than the global fermion parity, acting on a one-dimensional Majorana chain. Our study includes both, Gaussian circuits monitored with Born-rule measurements
as well as those with forced measurements.
Both types of Gaussian circuits belong to AZ symmetry class DIII according to the correspondence established in Ref.~\onlinecite{Jian2020}. The purpose of our study is the identification of the universality classes
of these two entanglement transitions and, in particular, to provide an answer to the question of whether or not they are in the same universality class. The latter question is in part motivated by the  corresponding question for 
entanglement transitions in monitored (non-Gaussian) many-body circuits of qudits with Haar-random
unitary gates where this question is not fully resolved to date, but for which conjectures exist \cite{Nahum2021AlltoAll}, as well as by the fact that for 
monitored Clifford circuits
(`stabilizer circuits'),  Born-rule and forced measurements yield  the same universality class of the  transition \cite{LiVasseurFisherLudwig2021}.

Our numerical simulations identified measurement-induced entanglement phase transitions between an area-law phase (with area-law entanglement scaling) and a critical phase (with logarithmic entanglement scaling) for both types of circuits (with Born-rule measurements and with forced measurements). Critical exponents at these entanglement transitions were numerically extracted (summarized in Table \ref{Tab:exponent_summary}). The value of the correlation length exponent $\nu$ at the entanglement transition
 induced by Born-rule measurements is found to be close to that at the forced-measurement-induced entanglement transition. 
 Both types of entanglement transitions turn out to exhibit two different sets of exponents associated with 
 rich scaling behavior of the (squared) Majorana-fermion correlation function $G$ 
in the long time limit of the circuit.
 For both entanglement transitions, the decay exponent $X_1$ for the averaged (squared) correlation function $\overline{G}$ was found to be very  different from the exponent $X_{typ}$ associated with the  decay of the typical correlation function, a signature of so-called multifractal scaling behavior of $G$. (See Appendix \ref{LabelSectionAppendixMultifractal}
 for a review, and Refs.~\onlinecite{LudwigHierarchies1990,Pixley2022Multifractality}.)
 Most importantly,
 the values of the decay exponent $X_{typ}$ of the {\it typical} correlation function were found to be widely different between the two types of entanglement transitions we studied. Furthermore, while the logarithm of the correlation function, $\log G$, is self-averaging for large separations (its average gives rise to the typical exponent $X_{typ}$), statistical
 fluctuations of $\log G$ about its average give rise to another 
universal quantity $x^{(2)}$
 which we also found to be widely different between the two types of entanglement transitions. These results provide strong evidence that the universality class of the Born-rule-measurement-induced entanglement transition is different from that of the forced-measurement-induced entanglement transition. Different values of the prefactor of the
 logarithmic dependence on subsystem size of the EE were also found for the two transitions.
 (See again the summary in Table \ref{Tab:exponent_summary}.)

 Moreover, we  provided a theoretical understanding for the numerically observed different universal behavior of these two entanglement transitions by identifying the underlying statistical-mechanics model describing
 the monitored Gaussian circuits with Born-rule and forced measurements discussed in this paper. Our ability to identify 
 the statistical-mechanics model originates from the correspondence established in Ref.~\onlinecite{Jian2020}. The crucial point is that, as already mentioned above,  {\it both} monitored Gaussian circuits
 in (1+1)-dimensional space-time dimensions, those
 with Born-rule and those with forced measurements, are described by the {\it same} AZ symmetry class.
In particular, based on the correspondence established in Ref. \onlinecite{Jian2020}, the forced-measurement-induced entanglement transition of the monitored Gaussian circuit in (1+1)-dimensional spacetime is identical to the thermal-metal-to-insulator transition in a two-spatial-dimensional non-interacting symmetry-class-DIII fermion system with a static disordered Hamiltonian. The latter is 
an Anderson localization transition. 
But even though the Born-rule-measurement-induced entanglement transition belongs to the same symmetry class DIII,  this transition is in a new universality class beyond Anderson localization transitions: Their common symmetry class DIII dictates that the behavior of both types of circuits  
is described (after averaging over all quantum trajectories)
by a two-dimensional principal chiral NLSM with a target space $\SO(n)$
(as a result of applying the replica trick). Such an NLSM has global $\SO(n)\times \SO(n)$ symmetry.
The replica limit $n\rightarrow 0$ corresponds to the monitored Gaussian circuit with forced measurement, while the replica limit $n\rightarrow 1$ corresponds to the monitored Gaussian circuit with Born-rule measurements. 
Our numerical results imply that the 
entanglement transition in the circuit with Born-rule measurements is thus in  a 
novel
universality class, different from the Anderson localization transition in the corresponding symmetry class DIII which
corresponds to the $n\to 0$ limit and forced measurements.
In either limit, the entanglement transition can be understood as a transition driven by 
proliferation of topological defects classified by the fundamental group of the target space,  which is  $\pi_1(\SO(n)) = \mathbb{Z}_2$ (for a generic $n$).
However, the different replica limits $n\rightarrow 0$ and $n\rightarrow 1$ will result in different universality classes for the two types of entanglement transitions. Both entanglement transitions, with Born-rule and with forced measurements, can be accessed by a perturbative RG treatment controlled (at least  in low order) by the small parameter $0 < \epsilon = 2-n$, which parallels the RG treatment by Fu and Kane in 
Ref.~\onlinecite{FuKane2012Sympletic} of the disorder-driven metal-insulator transition in the two-dimensional symplectic symmetry class AII. (Technical details of the generalization to the current symmetry class DIII, as well as an alternative formulation using supersymmetry which avoids the replica limit, will be presented  in separate follow-up work.)

We finally compared the entanglement transitions in the monitored Gaussian circuits discussed in the present paper with that in the circuits based on the loop model with crossings
\cite{Nahum2013LoopWithCrossing,NahumSkinnerMajDefect2020FF},
which is a highly fine-tuned version of the monitored Gaussian circuit of Majorana fermions. We provided numerical and analytical evidence showing that the
entanglement transition in the loop-model-based circuit
is in a universality class different from both entanglement transitions discussed in the present paper. As for the numerical argument, we
observed that the correlation length exponents obtained for the entanglement transitions in the generic Gaussian circuits, 
both for Born-rule measurements and for forced measurements,
differ significantly from the value for this exponent for the entanglement transition in the loop-model-based circuit (which can be simulated
as a classical loop model with crossings \cite{Nahum2013LoopWithCrossing}).
The obtained values of
the prefactor 
for the logarithmic dependence of the EE on
the subsystem size also appear different within error bars at the entanglement transitions.
As for the analytical argument, we noted that there is a manifest difference in symmetry: The statistical mechanics model for the entanglement transition in the generic Gaussian circuits discussed in the present paper possesses global $\SO(n)\times \SO(n)$ symmetry
in the replica limits $n\to 1$ and $n\to 0$, for Born-rule measurements and forced measurements, respectively, while that of
the loop-model-based circuit, on the other
hand, has only the smaller global $\SO(n)$ symmetry
(in the replica limit $n\to 1$).
Furthermore, we noted that corresponding statements can also be formulated within the supersymmetry approach in which no replica limit is taken. Given the different symmetries, one does not expect a relationship
between the entanglement transitions in the circuits discussed in this paper and that in the loop-model-based circuit.

Entanglement transitions in monitored Gaussian circuits are expected to be more tractable than those in general monitored circuits 
acting on qubits (qudits) (which are ``interacting"),
and can thus provide another angle into the nature of entanglement transitions.
The present paper demonstrates that the framework for classifying  non-unitary Gaussian circuits 
based on the ten-fold AZ symmetry classification which was developed in Ref.~\onlinecite{Jian2020} 
provides a concrete tool to successfully
identify measurement-induced entanglement transitions in Gaussian circuits monitored by Born-rule measurements and those monitored by forced measurements as well.
While the monitored Gaussian circuits with Born-rule 
and
forced measurements can be formulated (in any dimension) 
within
the same framework of the AZ symmetry classification as the Anderson localization problems, their universal behavior can  nevertheless be different:
It was shown in Ref.~\onlinecite{Jian2020} 
that entanglement transitions in monitored Gaussian circuits subject to forced measurements exactly correspond to Anderson localization transitions. As demonstrated in the present paper, 
the entanglement transition in a monitored Gaussian circuit subject to
Born-rule measurements can, while in the same AZ symmetry class, be in a 
novel
universality class that does not 
arise within the context of Anderson localization.
The framework developed in Ref.~\onlinecite{Jian2020}  
thus provides an approach to systematically investigate the appearance of such novel universality classes that can uniquely arise from entanglement transitions in Gaussian circuits monitored by Born-rule measurements.  
Understanding the differences between Born-rule and forced measurements will provide insight into novel
universal critical behavior that is unique to the context
of monitored random circuits and this will be investigated  
in follow-up works in other AZ symmetry classes and dimensions.

{\it Note added:} After the work on the present paper was completed, 
and after a summary of our results
had already been presented at the 2022 March Meeting of the American Physical Society \cite{HassanAPSabstract}, 
a paper (arXiv: 2210.05681) \cite{Lukasz2022} with some overlap with our numerics 
appeared on the arXiv while our work was being written up.

\acknowledgements
C.-M. J. thanks Haining Pan for helpful discussions on the numerical simulations of the monitored Gaussian circuits. This research is supported in part by a faculty
startup grant at Cornell University (C.-M.J.).

\appendix

\section{Covariance matrix formulation of Gaussian states and their evolution}
\label{app:covariance_matrix}
The numerical simulations of the monitored Gaussian circuits presented in this paper are carried out using the covariance matrix formulation reviewed in this appendix. 

In a system with $L$ Majorana fermion modes $\hgamma_{i=1,2,..., L}$, any Gaussian state, namely any state of non-interacting fermions, $|\Gamma\rangle$ can be fully captured by its covariance matrix 
\begin{align}
    \Gamma_{ij} = \Big\langle \frac{\i}{2} [\hgamma_i,\hgamma_j]\Big\rangle,
\end{align}
which encodes all the two-point Majorana fermion correlation functions. $\Gamma$ is a real anti-symmetric matrix that squares to $-\openone$, namely $\Gamma^2 = -\openone$, due to the purity of the Gaussian state $|\Gamma\rangle$. All muti-point correlation functions of a Gaussian state can be obtained from the two-point correlation functions via Wick's theorem. 

In the monitored Gaussian circuit acting on a one-dimensional Majorana chain, we can study the quantum dynamics of the system by calculating the evolution of the covariance matrix. For example, a Gaussian state $|\Gamma\rangle$ evolves in to the Gaussian state
\begin{align}
    |\Gamma'\rangle \equiv \frac{K_{(i, i+1)}(\vec{n})|\Gamma\rangle}{||K_{(i, i+1)}(\vec{n})|\Gamma\rangle||}
\end{align} under the action of the Kraus operator $K_{(i, i+1)}(\vec{n})$, defined in Eq. \eqref{LabelEqKvecn}, acting on the $i$th and $(i+1)$th sites of the Majorana chain. Following Ref. \onlinecite{Bravyi2004FermionicOptics}, the covariance matrix of the Gaussian state $|\Gamma'\rangle$ is given by 

\begin{widetext}
\begin{align*}
  \Gamma'& = \left( \begin{array}{cccc}
        \Gamma_{[1,i-1],[1,i-1]} & 0 &  \Gamma_{[1,i-1],[i+2,L]}\\
       0 &  -\i n_1 \sigma^y & 0\\
        \Gamma_{[i+2,L],[1,i-1]}  & 0 &  \Gamma_{[i+2,L], [i+2,L]}
    \end{array}  
    \right)
   \nonumber \\
   & -   
    \left( \begin{array}{cc}
        \Gamma_{[1,i-1],[i,i+1]}   & 0 \\
      0  &    -n_2 \openone +\i n_3 \sigma^y \\
          \Gamma_{[i+2,L],[i,i+1]}   &  0 
    \end{array}  
    \right).
       \left( \begin{array}{cc}
        \Gamma_{[i,i+1],[i,i+1]} &  \openone \\
        - \openone &   \i n_1 \sigma^y \\
    \end{array}  
    \right)^{-1}.
     \left( \begin{array}{ccc}
        \Gamma_{[i,i+1],[1,i-1]}   & 0 &   \Gamma_{[i,i+1],[i+2,L]} \\
      0  &    n_2 \openone +\i n_3 \sigma^y  & 0
    \end{array}  
    \right),
\end{align*}
\end{widetext}
where $\Gamma_{[j_1,j_2], [j_1',j_2']}$ represents the block of the matrix $\Gamma$ with the rows ranging from $j_1$ to $j_2$ and the columns ranging from $j_1'$ to $j_2'$. $\sigma^{x,y,z}$ are the Pauli matrices. Recall that $\Vec{n}=(n_1,n_2,n_3)$. 

Given the covariance matrix $\Gamma$ of the Gaussian state $|\Gamma\rangle$, one can directly calculate the subsystem von Neumann EE. For example, for a subsystem that is an interval starting from the $i$th site and ending on the $j$th site, the subsystem EE, is given by
\begin{align}
S_{[i,j]} = -\frac{1}{2 }\sum_{s=\pm 1}\Tr\left( \frac{\openone + s \i \ \Gamma_{[i,j], [i,j]} }{2} \log \frac{\openone + s \i \ \Gamma_{[i,j], [i,j]} }{2}\right).   
\end{align}

\section{Monte Carlo sampling for monitored Gaussian circuits with Born-rule measurements}
\label{LabelSectionAppendixMonteCarloSimulations}

In the monitored Gaussian circuits circuit with Born-rule measurement, for a given pair of sites $(i,i+1)$ and at a given time step, the Born-rule probability density for the occurrence of the Kraus operator $K(\Vec{n})$, associated with the measurement outcome labeled by $\Vec{n}$, depends on the state $|\Gamma\rangle$ of the system at that time step (before the action of $K(\Vec{n})$ is implemented). As explained in Eq. \eqref{Eq:Born-rule_average}, this probability density is given by 
\begin{align}
 \Pb(s,\phi)\ ds d\phi \equiv \frac{1}{2\pi } p(s) \Tilde{w}_Y(s)   \langle K^\dag(\vec{n}_Y) K(\vec{n}_Y)\rangle \ ds d\phi 
\end{align}
with $Y=A$ or $B$ depending on which sublattice the Kraus operator belongs to in the circuit geometry. $p(s)$ is given in Eq. \eqref{eq:p-dist}. The $\vec{n}_Y$ is parameterized by $(s,\varphi)$ following Eq. \eqref{eq:nAB_reparameterized}:
\begin{align}
& \vec{n}_A = (s,\sqrt{1-s^2}\cos\varphi,\sqrt{1-s^2}\sin\varphi), \nonumber \\
& \vec{n}_B = (\sqrt{1-s^2}\sin\varphi,s,\sqrt{1-s^2}\cos\varphi).     
\end{align}
$\langle K^\dag(\vec{n}_Y) K(\vec{n}_Y) \rangle$ is evaluated with respect to the state $|\Gamma\rangle$ which can be calculated using the covariance matrix (introduced in App. \ref{app:covariance_matrix})

In a numerical implementation of the monitored Gaussian circuit with Born-rule measurements, it is not practical to sample cover all possible values of $(s,\varphi)$. To determine which Kraus operator $K(\Vec{n})$ to apply, we use the Metropolis algorithm as an importance sampling scheme:
\begin{algorithm}[H] \caption{Importance sampling of measurement outcome $\Vec{n}_Y$ according to the Born-rule probability} \label{alg1}
\begin{algorithmic}[1]
\State  To initialize, pick a random $n_Y^{(0)} \in S^2$ by picking a random pair of variables $s^{(0)}\in [-1,1]$ and $\phi\in [0,2\pi)$, and compute the Born-rule probability density $\textsc{P}(s^{(0)},\phi^{(0)})$ with respect to the input state  $|\Gamma\rangle$.
\For{$n$ in $N_\text{iter}$}
\State Propose a new pair $(s^{(n)},\varphi^{(n)})$, i.e. a new $n_Y^{(n)}$.
\State Compute $\Pb(s^{(n)},\varphi^{(n)})$ and accept the Monte Carlo update with probability
 \begin{align*}
 & p_\text{acc}((s^{(n-1)},\varphi^{(n-1)}) \to (s^{(n)},\varphi^{(n)}) )  \\ 
 & ~~~~~~~~~~~~~~~~~~~~~~~ =
\min \left( 1, \frac{\Pb(s^{(n)},\varphi^{(n)})}{\Pb(s^{(n-1)},\varphi^{(n-1)})}\right).
\end{align*}
    \EndFor
    \end{algorithmic}
\end{algorithm}
\noindent Here, $N_\text{iter}$ is the number of iterations. 

This Metropolis algorithm produces the correct Born-rule probability distribution of $(s,\varphi)$ due to detailed balance:
\begin{align}
   & \Pb(s,\varphi) \ p_\text{acc}((s,\varphi) \to (s',\varphi') ) 
    \nonumber\\
   & ~~~~~~~~~~~~ = \Pb(s',\varphi')
    \ p_\text{acc}((s',\varphi') \to (s,\varphi) )
\end{align}
which we verify numerically.

\section{Method to extract the correlation length exponent $\nu$ for the entanglement transitions}

\label{app:nu_extraction}

In Sec. \ref{LabelSectionNumericalResults}, we determine the entanglement phase transition points and the associated correlation length critical exponents $\nu$ using the scaling collapse of the two-interval mutual information in a configuration with a fixed cross-ratio $\eta = \sin^2 (\pi/8)$. 

Take the case of monitored Gaussian circuits with Born-rule measurements as an example. At fixed $b_2$ and $\eta$, the two-interval mutual information $I$ is treated as a function of the parameter $a_2$ and the total system size $L$. Our objective is to determine the value of $a_2$ at the entanglement transition, denoted as $a_{2c}$, and the corresponding correlation length exponent $\nu$. Around the transition, a scaling collapse is expected such that $|I(a_c, L) - I(a_{2c},L)|$ is a function that depends only on the single variable $(a-a_{2c})L^{1/\nu}$. 

We apply the algorithm introduced in Ref. \onlinecite{Skinner2019MIPT} to obtain the ``optimal" values of $a_{2c}$ and $\nu$. 
The algorithm objective is to minimize a cost function $R(a_{2c}, \nu)$ which 
essentially measures the deviation from a data collapse on a universal (unknown) curve. The corresponding optimal values are our estimates for $a_{2c}$ and $\nu$. For a given value of $a_{2c}$ and $\nu$, we estimate $I(a_{2c}, L)$ for each system size $L$ by a piece-wise linear interpolation. We then calculate the function $y_L (x)= I(a_2, L) - I(a_{2c}, L)$ and  $x=(a_2-a_{2c})L^{1/\nu}$ for given values of $a_2$ and $L$ from the data set. This gives a family of curves $y_L(x)$ vs.~$x$, which we wish to collapse on a single curve. Next, we sample from these curves at a discrete set of points $\{ x_i \}$ and define the cost function
\begin{align}
R = \sum_{i, L} \left[ y_L(x_i) - \bar{y}(x_i) \right]^2,
\end{align}
in terms of sum of the variance of $y_L(x_i)$ for different system sizes, where
\begin{align}
    \bar{y}(x_i) = \frac{1}{N_L} \sum_L y_L(x_i),
\end{align}
is the mean value at point $x_i$, and $N_L$ is the number of different system sizes in the data set. We should note that again we use piece-wise linear interpolation to estimate $y_L(x_i)$ and omit a point for a given $L$ if it is outside the range of the data set for that particular $L$.
Finally,  we search numerically for the values of $a_{2c}$ and $\nu$ that minimize the objective function.

\begin{figure}
    \centering
    \includegraphics[scale=0.8]{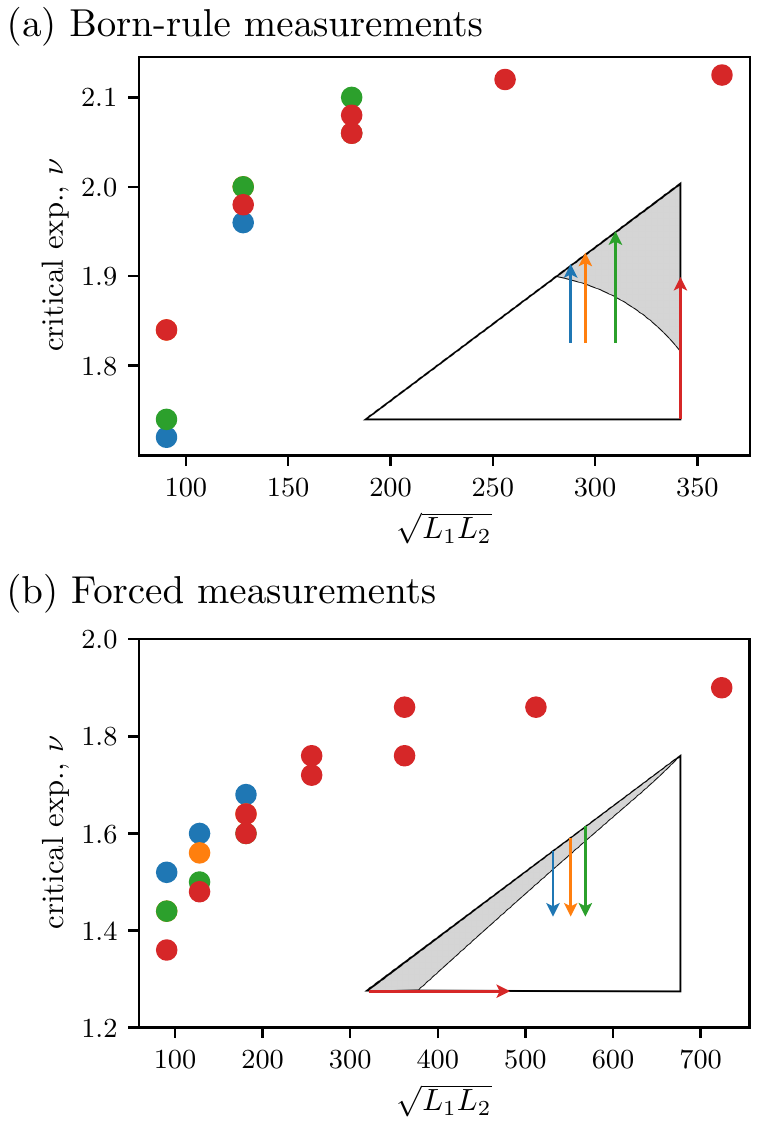}
    \caption{Finite-size effects in evaluating the critical exponent $\nu$ associated with the mutual information. In panel (a), $a_2$ is swept for a fixed value of $b_2$ as follows: $b_2=0.66$ (blue), $0.7$ (orange), $0.8$ (green), $1.0$ (red, corresponding to the transition discussed in the main text). Similarly, in panel (b), we sweep $a_2$ for $b_2=0.6$ (blue), $0.66$ (orange), and $0.7$ (green), while red points correspond to sweeping $b_2$ for $a_2=0$ as in the transition investigated in the main text.}
    \label{fig:fs-exp}
\end{figure}

In order to estimate the uncertainty in our results for $a_{2c}$ and $\nu$, we examine how the optimum point change when we run the minimization algorithm on a subset of data. In particular, we choose every pair of system sizes (call them $L_1$ and $L_2$) and calculate the optimum values of $a_{2c}$ and $\nu$ and use their variation as a proxy for the uncertainty. The resulting values of $\nu$ for various cuts in the phase diagram are plotted in Fig.~\ref{fig:fs-exp}. We observe that there is a finite-size effect which leads to smaller $\nu$'s for smaller system sizes.  However, there is a small variation among different transition points for a  given pair of system sizes. Note that the values of $\nu$ reported in the main text were obtained from the optimization process, including all system sizes corresponding to the paths shown in red in this figure.

\section{Multifractal fermion correlation function}
\label{LabelSectionAppendixMultifractal}

On a length-$L$ Majorana chain with periodic boundary condition,
we consider the square of the Majorana fermion 2-point function at lattice sites
$p$ and $p+r$ on the final time slice of a fixed realization ${\cal C}$ of the entire spacetime of the monitored Gaussian circuit (associated with a certain quantum trajectory)
\begin{eqnarray}
\label{LabelEqDEFCorrelator}
G(p, p+r; {\cal C})
=
\left(
\langle {\bf i}
{\hat \gamma}_p {\hat \gamma}_{p+r}\rangle_{\cal C}
\right)^2.
\end{eqnarray}
At the transition between the critical and the area-law phase (with Born-rule or forced measurements),
the $N$-th 
moments of this correlation function will scale as
\begin{eqnarray}
\label{LabelEqScalingOfMomentsOfCorrelator}
&&
\overline{{\left[
G(p, p+r;{\cal C}
\right]}^N}
\sim
\frac{B_N}{R(r)^{2X_N}}.
\end{eqnarray}
Here, $R(r)$ is the chord distance
\begin{align}
    R(r) \equiv 
{L\over \pi} \sin\left({\pi\over L} |r|
\right),
\end{align}
where $L$ is the spatial size of the chain (with periodic boundary conditions). The overbar $\overline{~\cdots~}$ represents the  averaging over all circuit realization $\cal{C}$ according to the probability distribution given by the Born-rule or forced measurements (as explained in Sec. \ref{LabelSectionMonitoredGaussianCircuits} ).
In general, when the exponents
$X_N$ are not linear functions of the moment order $N$, the correlation function is 
referred to as being 
``multifractal''~\cite{LudwigHierarchies1990,Pixley2022Multifractality,DuplantierLudwigPRL1991}\footnote{When all 
moments are identical, one may not  want to refer to this as ``multifractal'', as this is in a way a ``trivial'' case where there is really only one moment, namely the 1st moment.}. (The $N$-dependence of the amplitudes $B_N$ does not describe universal properties, and we will not be concerned with this dependence.)

We now use the standard cumulant expansion on the left hand side of Eq. \eqref{LabelEqScalingOfMomentsOfCorrelator}, 
\begin{eqnarray}
\label{LabelEqDEFCumulantsOfCorrelator}
&&
\overline{{\left[
G(p, p+r;{\cal C}
\right]}^N}
=
\\ \nonumber
&&=
\exp\left\{N ~\overline{\log G}
+{N^2\over 2!}
\left[~\overline{\left (\log G\right )^2}
-
\overline{\left (\log G\right )}^2
\right ] + ...
\right\}
\qquad \quad
\\ \nonumber
&& \equiv
\exp\left\{N ~\kappa_1
+{N^2\over 2!}\kappa_2 + {N^3\over 3!} \kappa_3 + ...
\right\}
\end{eqnarray}
where $\kappa_j$ denotes the $j$-th cumulant of the random variable $\log G(p, p+r; {\cal C})$,
\begin{eqnarray}
\label{LabelEqCumulants}
&\kappa_1(r) & = \overline{\left[\log G(p, p+r;{\cal C})\right]},
 \nonumber \\
&\kappa_2 (r)&=\overline{{\left[\log G(p, p+r;{\cal C})\right]}^2}
-
\overline{\left[\log G(p, p+r;{\cal C})\right]}^2.
\nonumber \\
& & \hspace{2.5cm}  \vdots 
\end{eqnarray}
Comparing this with the Taylor expansion in the (moment order) $N$-dependence of the exponents $X_N$,
\begin{eqnarray}
\label{LabelEqExpansionOfXN}
X_N =
N x^{(1)} + {N^2\over 2!} x^{(2)}+ {N^3\over 3!} x^{(3)}+
...
\end{eqnarray}
used on the right-hand side of the same equation,  Eq.~\eqref{LabelEqScalingOfMomentsOfCorrelator}, we conclude, owing to the scaling in
Eq. (\ref{LabelEqScalingOfMomentsOfCorrelator}) of all moments of the correlation function, that
{\it all cumulants $\kappa_j$ must scale linearly with $\log R(r)$}, where the coefficients of proportionality are universal and are nothing but  the Taylor coefficients $x^{(j)}$ appearing in
the Taylor expansion of the critical exponents $X_N$ in the moment order, Eq.~(\ref{LabelEqExpansionOfXN}):
\begin{eqnarray}
\label{LabelEqCoefficientsOfCumulants}
\kappa_j = -  2 x^{(j)}~ \log R(r).
\end{eqnarray}
Here $x^{(j=1)}$ is conventionally referred to as 
the ``typical" critical exponent,
$X_{typ} := x^{(j=1)}$.
We re-iterate  that the coefficients of  proportionality of the cumulants versus $\log R(r)$ are universal properties of the universality class of the phase transition considered.
In other words,   {\it two transitions exhibiting different values of these coefficients $x^{(j)}$ must be in different universality classes}.

In the present paper, we numerically extract the coefficients $x^{(1)}$ and $x^{(2)}$ (from  the 1st and the 2nd cumulants), and show that they are significantly different
for the entanglement transitions in the monitored Gaussian circuits with 
Born-rule measurements and those with forced measurements. (See Table \ref{Tab:exponent_summary}.) This  will serve as strong evidence that these two transitions are in different universality classes. We also note that since $x^{(2)}$ is not vanishing for both transitions, they both exhibit multifractal correlations on the final time slice (since the dependence of the
exponents in Eq.~\eqref{LabelEqExpansionOfXN}
on the moment order
$N$ will then be non-linear).

\newpage

\bibliography{refs.bib}

\begin{thebibliography}{77}%
\makeatletter
\providecommand \@ifxundefined [1]{%
 \@ifx{#1\undefined}
}%
\providecommand \@ifnum [1]{%
 \ifnum #1\expandafter \@firstoftwo
 \else \expandafter \@secondoftwo
 \fi
}%
\providecommand \@ifx [1]{%
 \ifx #1\expandafter \@firstoftwo
 \else \expandafter \@secondoftwo
 \fi
}%
\providecommand \natexlab [1]{#1}%
\providecommand \enquote  [1]{``#1''}%
\providecommand \bibnamefont  [1]{#1}%
\providecommand \bibfnamefont [1]{#1}%
\providecommand \citenamefont [1]{#1}%
\providecommand \href@noop [0]{\@secondoftwo}%
\providecommand \href [0]{\begingroup \@sanitize@url \@href}%
\providecommand \@href[1]{\@@startlink{#1}\@@href}%
\providecommand \@@href[1]{\endgroup#1\@@endlink}%
\providecommand \@sanitize@url [0]{\catcode `\\12\catcode `\$12\catcode
  `\&12\catcode `\#12\catcode `\^12\catcode `\_12\catcode `\%12\relax}%
\providecommand \@@startlink[1]{}%
\providecommand \@@endlink[0]{}%
\providecommand \url  [0]{\begingroup\@sanitize@url \@url }%
\providecommand \@url [1]{\endgroup\@href {#1}{\urlprefix }}%
\providecommand \urlprefix  [0]{URL }%
\providecommand \Eprint [0]{\href }%
\providecommand \doibase [0]{http://dx.doi.org/}%
\providecommand \selectlanguage [0]{\@gobble}%
\providecommand \bibinfo  [0]{\@secondoftwo}%
\providecommand \bibfield  [0]{\@secondoftwo}%
\providecommand \translation [1]{[#1]}%
\providecommand \BibitemOpen [0]{}%
\providecommand \bibitemStop [0]{}%
\providecommand \bibitemNoStop [0]{.\EOS\space}%
\providecommand \EOS [0]{\spacefactor3000\relax}%
\providecommand \BibitemShut  [1]{\csname bibitem#1\endcsname}%
\let\auto@bib@innerbib\@empty
\bibitem [{\citenamefont {{Li}}\ \emph {et~al.}(2018)\citenamefont {{Li}},
  \citenamefont {{Chen}},\ and\ \citenamefont
  {{Fisher}}}]{LiChenFisher2018MIPT}%
  \BibitemOpen
  \bibfield  {author} {\bibinfo {author} {\bibfnamefont {Y.}~\bibnamefont
  {{Li}}}, \bibinfo {author} {\bibfnamefont {X.}~\bibnamefont {{Chen}}}, \ and\
  \bibinfo {author} {\bibfnamefont {M.~P.~A.}\ \bibnamefont {{Fisher}}},\
  }\href {\doibase 10.1103/PhysRevB.98.205136} {\bibfield  {journal} {\bibinfo
  {journal} {\prb}\ }\textbf {\bibinfo {volume} {98}},\ \bibinfo {eid} {205136}
  (\bibinfo {year} {2018})},\ \Eprint {http://arxiv.org/abs/1808.06134}
  {arXiv:1808.06134 [quant-ph]} \BibitemShut {NoStop}%
\bibitem [{\citenamefont {{Skinner}}\ \emph {et~al.}(2019)\citenamefont
  {{Skinner}}, \citenamefont {{Ruhman}},\ and\ \citenamefont
  {{Nahum}}}]{Skinner2019MIPT}%
  \BibitemOpen
  \bibfield  {author} {\bibinfo {author} {\bibfnamefont {B.}~\bibnamefont
  {{Skinner}}}, \bibinfo {author} {\bibfnamefont {J.}~\bibnamefont {{Ruhman}}},
  \ and\ \bibinfo {author} {\bibfnamefont {A.}~\bibnamefont {{Nahum}}},\ }\href
  {\doibase 10.1103/PhysRevX.9.031009} {\bibfield  {journal} {\bibinfo
  {journal} {Physical Review X}\ }\textbf {\bibinfo {volume} {9}},\ \bibinfo
  {eid} {031009} (\bibinfo {year} {2019})},\ \Eprint
  {http://arxiv.org/abs/1808.05953} {arXiv:1808.05953 [cond-mat.stat-mech]}
  \BibitemShut {NoStop}%
\bibitem [{\citenamefont {{Chan}}\ \emph {et~al.}(2019)\citenamefont {{Chan}},
  \citenamefont {{Nandkishore}}, \citenamefont {{Pretko}},\ and\ \citenamefont
  {{Smith}}}]{Chan2019MIPT}%
  \BibitemOpen
  \bibfield  {author} {\bibinfo {author} {\bibfnamefont {A.}~\bibnamefont
  {{Chan}}}, \bibinfo {author} {\bibfnamefont {R.~M.}\ \bibnamefont
  {{Nandkishore}}}, \bibinfo {author} {\bibfnamefont {M.}~\bibnamefont
  {{Pretko}}}, \ and\ \bibinfo {author} {\bibfnamefont {G.}~\bibnamefont
  {{Smith}}},\ }\href {\doibase 10.1103/PhysRevB.99.224307} {\bibfield
  {journal} {\bibinfo  {journal} {\prb}\ }\textbf {\bibinfo {volume} {99}},\
  \bibinfo {eid} {224307} (\bibinfo {year} {2019})},\ \Eprint
  {http://arxiv.org/abs/1808.05949} {arXiv:1808.05949 [cond-mat.stat-mech]}
  \BibitemShut {NoStop}%
\bibitem [{\citenamefont {{Choi}}\ \emph {et~al.}(2020)\citenamefont {{Choi}},
  \citenamefont {{Bao}}, \citenamefont {{Qi}},\ and\ \citenamefont
  {{Altman}}}]{ChoiBaoQiEhud2020MIPT}%
  \BibitemOpen
  \bibfield  {author} {\bibinfo {author} {\bibfnamefont {S.}~\bibnamefont
  {{Choi}}}, \bibinfo {author} {\bibfnamefont {Y.}~\bibnamefont {{Bao}}},
  \bibinfo {author} {\bibfnamefont {X.-L.}\ \bibnamefont {{Qi}}}, \ and\
  \bibinfo {author} {\bibfnamefont {E.}~\bibnamefont {{Altman}}},\ }\href
  {\doibase 10.1103/PhysRevLett.125.030505} {\bibfield  {journal} {\bibinfo
  {journal} {\prl}\ }\textbf {\bibinfo {volume} {125}},\ \bibinfo {eid}
  {030505} (\bibinfo {year} {2020})},\ \Eprint
  {http://arxiv.org/abs/1903.05124} {arXiv:1903.05124 [quant-ph]} \BibitemShut
  {NoStop}%
\bibitem [{\citenamefont {{Gullans}}\ and\ \citenamefont
  {{Huse}}(2020)}]{GullansHuse2020Purification}%
  \BibitemOpen
  \bibfield  {author} {\bibinfo {author} {\bibfnamefont {M.~J.}\ \bibnamefont
  {{Gullans}}}\ and\ \bibinfo {author} {\bibfnamefont {D.~A.}\ \bibnamefont
  {{Huse}}},\ }\href {\doibase 10.1103/PhysRevX.10.041020} {\bibfield
  {journal} {\bibinfo  {journal} {Physical Review X}\ }\textbf {\bibinfo
  {volume} {10}},\ \bibinfo {eid} {041020} (\bibinfo {year} {2020})},\ \Eprint
  {http://arxiv.org/abs/1905.05195} {arXiv:1905.05195 [quant-ph]} \BibitemShut
  {NoStop}%
\bibitem [{\citenamefont {Gullans}\ and\ \citenamefont
  {Huse}(2020)}]{Guallans2020ScalableProbes}%
  \BibitemOpen
  \bibfield  {author} {\bibinfo {author} {\bibfnamefont {M.~J.}\ \bibnamefont
  {Gullans}}\ and\ \bibinfo {author} {\bibfnamefont {D.~A.}\ \bibnamefont
  {Huse}},\ }\href {\doibase 10.1103/PhysRevLett.125.070606} {\bibfield
  {journal} {\bibinfo  {journal} {Phys. Rev. Lett.}\ }\textbf {\bibinfo
  {volume} {125}},\ \bibinfo {pages} {070606} (\bibinfo {year}
  {2020})}\BibitemShut {NoStop}%
\bibitem [{\citenamefont {{Jian}}\ \emph {et~al.}(2020)\citenamefont {{Jian}},
  \citenamefont {{You}}, \citenamefont {{Vasseur}},\ and\ \citenamefont
  {{Ludwig}}}]{JianYouVassuerLudwig2020MIPT}%
  \BibitemOpen
  \bibfield  {author} {\bibinfo {author} {\bibfnamefont {C.-M.}\ \bibnamefont
  {{Jian}}}, \bibinfo {author} {\bibfnamefont {Y.-Z.}\ \bibnamefont {{You}}},
  \bibinfo {author} {\bibfnamefont {R.}~\bibnamefont {{Vasseur}}}, \ and\
  \bibinfo {author} {\bibfnamefont {A.~W.~W.}\ \bibnamefont {{Ludwig}}},\
  }\href {\doibase 10.1103/PhysRevB.101.104302} {\bibfield  {journal} {\bibinfo
   {journal} {\prb}\ }\textbf {\bibinfo {volume} {101}},\ \bibinfo {eid}
  {104302} (\bibinfo {year} {2020})},\ \Eprint
  {http://arxiv.org/abs/1908.08051} {arXiv:1908.08051 [cond-mat.stat-mech]}
  \BibitemShut {NoStop}%
\bibitem [{\citenamefont {{Bao}}\ \emph {et~al.}(2020)\citenamefont {{Bao}},
  \citenamefont {{Choi}},\ and\ \citenamefont {{Altman}}}]{BaoChoiAltman2020}%
  \BibitemOpen
  \bibfield  {author} {\bibinfo {author} {\bibfnamefont {Y.}~\bibnamefont
  {{Bao}}}, \bibinfo {author} {\bibfnamefont {S.}~\bibnamefont {{Choi}}}, \
  and\ \bibinfo {author} {\bibfnamefont {E.}~\bibnamefont {{Altman}}},\ }\href
  {\doibase 10.1103/PhysRevB.101.104301} {\bibfield  {journal} {\bibinfo
  {journal} {\prb}\ }\textbf {\bibinfo {volume} {101}},\ \bibinfo {eid}
  {104301} (\bibinfo {year} {2020})},\ \Eprint
  {http://arxiv.org/abs/1908.04305} {arXiv:1908.04305 [cond-mat.stat-mech]}
  \BibitemShut {NoStop}%
\bibitem [{\citenamefont {Li}\ \emph {et~al.}(2019)\citenamefont {Li},
  \citenamefont {Chen},\ and\ \citenamefont {Fisher}}]{LiChenFisher2019}%
  \BibitemOpen
  \bibfield  {author} {\bibinfo {author} {\bibfnamefont {Y.}~\bibnamefont
  {Li}}, \bibinfo {author} {\bibfnamefont {X.}~\bibnamefont {Chen}}, \ and\
  \bibinfo {author} {\bibfnamefont {M.~P.~A.}\ \bibnamefont {Fisher}},\ }\href
  {\doibase 10.1103/PhysRevB.100.134306} {\bibfield  {journal} {\bibinfo
  {journal} {Phys. Rev. B}\ }\textbf {\bibinfo {volume} {100}},\ \bibinfo
  {pages} {134306} (\bibinfo {year} {2019})}\BibitemShut {NoStop}%
\bibitem [{\citenamefont {{Szyniszewski}}\ \emph {et~al.}(2019)\citenamefont
  {{Szyniszewski}}, \citenamefont {{Romito}},\ and\ \citenamefont
  {{Schomerus}}}]{Szyniszewski2019}%
  \BibitemOpen
  \bibfield  {author} {\bibinfo {author} {\bibfnamefont {M.}~\bibnamefont
  {{Szyniszewski}}}, \bibinfo {author} {\bibfnamefont {A.}~\bibnamefont
  {{Romito}}}, \ and\ \bibinfo {author} {\bibfnamefont {H.}~\bibnamefont
  {{Schomerus}}},\ }\href {\doibase 10.1103/PhysRevB.100.064204} {\bibfield
  {journal} {\bibinfo  {journal} {\prb}\ }\textbf {\bibinfo {volume} {100}},\
  \bibinfo {eid} {064204} (\bibinfo {year} {2019})},\ \Eprint
  {http://arxiv.org/abs/1903.05452} {arXiv:1903.05452 [cond-mat.stat-mech]}
  \BibitemShut {NoStop}%
\bibitem [{\citenamefont {{Tang}}\ and\ \citenamefont
  {{Zhu}}(2020)}]{Tang2020nonintegrable}%
  \BibitemOpen
  \bibfield  {author} {\bibinfo {author} {\bibfnamefont {Q.}~\bibnamefont
  {{Tang}}}\ and\ \bibinfo {author} {\bibfnamefont {W.}~\bibnamefont {{Zhu}}},\
  }\href {\doibase 10.1103/PhysRevResearch.2.013022} {\bibfield  {journal}
  {\bibinfo  {journal} {Physical Review Research}\ }\textbf {\bibinfo {volume}
  {2}},\ \bibinfo {eid} {013022} (\bibinfo {year} {2020})},\ \Eprint
  {http://arxiv.org/abs/1908.11253} {arXiv:1908.11253 [cond-mat.stat-mech]}
  \BibitemShut {NoStop}%
\bibitem [{\citenamefont {{Lopez-Piqueres}}\ \emph {et~al.}(2020)\citenamefont
  {{Lopez-Piqueres}}, \citenamefont {{Ware}},\ and\ \citenamefont
  {{Vasseur}}}]{Vasseur2020TreeTN}%
  \BibitemOpen
  \bibfield  {author} {\bibinfo {author} {\bibfnamefont {J.}~\bibnamefont
  {{Lopez-Piqueres}}}, \bibinfo {author} {\bibfnamefont {B.}~\bibnamefont
  {{Ware}}}, \ and\ \bibinfo {author} {\bibfnamefont {R.}~\bibnamefont
  {{Vasseur}}},\ }\href {\doibase 10.1103/PhysRevB.102.064202} {\bibfield
  {journal} {\bibinfo  {journal} {\prb}\ }\textbf {\bibinfo {volume} {102}},\
  \bibinfo {eid} {064202} (\bibinfo {year} {2020})},\ \Eprint
  {http://arxiv.org/abs/2003.01138} {arXiv:2003.01138 [cond-mat.stat-mech]}
  \BibitemShut {NoStop}%
\bibitem [{\citenamefont {{Lavasani}}\ \emph {et~al.}(2021)\citenamefont
  {{Lavasani}}, \citenamefont {{Alavirad}},\ and\ \citenamefont
  {{Barkeshli}}}]{BarkeshiliTopoCircuit2021}%
  \BibitemOpen
  \bibfield  {author} {\bibinfo {author} {\bibfnamefont {A.}~\bibnamefont
  {{Lavasani}}}, \bibinfo {author} {\bibfnamefont {Y.}~\bibnamefont
  {{Alavirad}}}, \ and\ \bibinfo {author} {\bibfnamefont {M.}~\bibnamefont
  {{Barkeshli}}},\ }\href {\doibase 10.1038/s41567-020-01112-z} {\bibfield
  {journal} {\bibinfo  {journal} {Nature Physics}\ }\textbf {\bibinfo {volume}
  {17}},\ \bibinfo {pages} {342} (\bibinfo {year} {2021})},\ \Eprint
  {http://arxiv.org/abs/2004.07243} {arXiv:2004.07243 [quant-ph]} \BibitemShut
  {NoStop}%
\bibitem [{\citenamefont {{Sang}}\ and\ \citenamefont
  {{Hsieh}}(2021)}]{SangHsieh2021}%
  \BibitemOpen
  \bibfield  {author} {\bibinfo {author} {\bibfnamefont {S.}~\bibnamefont
  {{Sang}}}\ and\ \bibinfo {author} {\bibfnamefont {T.~H.}\ \bibnamefont
  {{Hsieh}}},\ }\href {\doibase 10.1103/PhysRevResearch.3.023200} {\bibfield
  {journal} {\bibinfo  {journal} {Physical Review Research}\ }\textbf {\bibinfo
  {volume} {3}},\ \bibinfo {eid} {023200} (\bibinfo {year} {2021})},\ \Eprint
  {http://arxiv.org/abs/2004.09509} {arXiv:2004.09509 [cond-mat.stat-mech]}
  \BibitemShut {NoStop}%
\bibitem [{\citenamefont {Sang}\ \emph {et~al.}(2021)\citenamefont {Sang},
  \citenamefont {Li}, \citenamefont {Zhou}, \citenamefont {Chen}, \citenamefont
  {Hsieh},\ and\ \citenamefont {Fisher}}]{Sang2021Negativity}%
  \BibitemOpen
  \bibfield  {author} {\bibinfo {author} {\bibfnamefont {S.}~\bibnamefont
  {Sang}}, \bibinfo {author} {\bibfnamefont {Y.}~\bibnamefont {Li}}, \bibinfo
  {author} {\bibfnamefont {T.}~\bibnamefont {Zhou}}, \bibinfo {author}
  {\bibfnamefont {X.}~\bibnamefont {Chen}}, \bibinfo {author} {\bibfnamefont
  {T.~H.}\ \bibnamefont {Hsieh}}, \ and\ \bibinfo {author} {\bibfnamefont
  {M.~P.}\ \bibnamefont {Fisher}},\ }\href {\doibase
  10.1103/PRXQuantum.2.030313} {\bibfield  {journal} {\bibinfo  {journal} {PRX
  Quantum}\ }\textbf {\bibinfo {volume} {2}},\ \bibinfo {pages} {030313}
  (\bibinfo {year} {2021})}\BibitemShut {NoStop}%
\bibitem [{\citenamefont {{Ippoliti}}\ \emph {et~al.}(2021)\citenamefont
  {{Ippoliti}}, \citenamefont {{Gullans}}, \citenamefont {{Gopalakrishnan}},
  \citenamefont {{Huse}},\ and\ \citenamefont
  {{Khemani}}}]{Ippoliti2021MeasurementOnly}%
  \BibitemOpen
  \bibfield  {author} {\bibinfo {author} {\bibfnamefont {M.}~\bibnamefont
  {{Ippoliti}}}, \bibinfo {author} {\bibfnamefont {M.~J.}\ \bibnamefont
  {{Gullans}}}, \bibinfo {author} {\bibfnamefont {S.}~\bibnamefont
  {{Gopalakrishnan}}}, \bibinfo {author} {\bibfnamefont {D.~A.}\ \bibnamefont
  {{Huse}}}, \ and\ \bibinfo {author} {\bibfnamefont {V.}~\bibnamefont
  {{Khemani}}},\ }\href {\doibase 10.1103/PhysRevX.11.011030} {\bibfield
  {journal} {\bibinfo  {journal} {Physical Review X}\ }\textbf {\bibinfo
  {volume} {11}},\ \bibinfo {eid} {011030} (\bibinfo {year} {2021})},\ \Eprint
  {http://arxiv.org/abs/2004.09560} {arXiv:2004.09560 [quant-ph]} \BibitemShut
  {NoStop}%
\bibitem [{\citenamefont {{Fuji}}\ and\ \citenamefont
  {{Ashida}}(2020)}]{FujiAshida2020}%
  \BibitemOpen
  \bibfield  {author} {\bibinfo {author} {\bibfnamefont {Y.}~\bibnamefont
  {{Fuji}}}\ and\ \bibinfo {author} {\bibfnamefont {Y.}~\bibnamefont
  {{Ashida}}},\ }\href {\doibase 10.1103/PhysRevB.102.054302} {\bibfield
  {journal} {\bibinfo  {journal} {\prb}\ }\textbf {\bibinfo {volume} {102}},\
  \bibinfo {eid} {054302} (\bibinfo {year} {2020})},\ \Eprint
  {http://arxiv.org/abs/2004.11957} {arXiv:2004.11957 [cond-mat.stat-mech]}
  \BibitemShut {NoStop}%
\bibitem [{\citenamefont {{Lunt}}\ and\ \citenamefont
  {{Pal}}(2020)}]{LuntPal2020}%
  \BibitemOpen
  \bibfield  {author} {\bibinfo {author} {\bibfnamefont {O.}~\bibnamefont
  {{Lunt}}}\ and\ \bibinfo {author} {\bibfnamefont {A.}~\bibnamefont {{Pal}}},\
  }\href {\doibase 10.1103/PhysRevResearch.2.043072} {\bibfield  {journal}
  {\bibinfo  {journal} {Physical Review Research}\ }\textbf {\bibinfo {volume}
  {2}},\ \bibinfo {eid} {043072} (\bibinfo {year} {2020})},\ \Eprint
  {http://arxiv.org/abs/2005.13603} {arXiv:2005.13603 [quant-ph]} \BibitemShut
  {NoStop}%
\bibitem [{\citenamefont {Lang}\ and\ \citenamefont
  {B\"uchler}(2020)}]{LangBuchler2020}%
  \BibitemOpen
  \bibfield  {author} {\bibinfo {author} {\bibfnamefont {N.}~\bibnamefont
  {Lang}}\ and\ \bibinfo {author} {\bibfnamefont {H.~P.}\ \bibnamefont
  {B\"uchler}},\ }\href {\doibase 10.1103/PhysRevB.102.094204} {\bibfield
  {journal} {\bibinfo  {journal} {Phys. Rev. B}\ }\textbf {\bibinfo {volume}
  {102}},\ \bibinfo {pages} {094204} (\bibinfo {year} {2020})}\BibitemShut
  {NoStop}%
\bibitem [{\citenamefont {{Vijay}}(2020)}]{Vijay2020VolumeLaw}%
  \BibitemOpen
  \bibfield  {author} {\bibinfo {author} {\bibfnamefont {S.}~\bibnamefont
  {{Vijay}}},\ }\href@noop {} {\bibfield  {journal} {\bibinfo  {journal} {arXiv
  e-prints}\ ,\ \bibinfo {eid} {arXiv:2005.03052}} (\bibinfo {year} {2020})},\
  \Eprint {http://arxiv.org/abs/2005.03052} {arXiv:2005.03052 [quant-ph]}
  \BibitemShut {NoStop}%
\bibitem [{\citenamefont {Nahum}\ \emph {et~al.}(2021)\citenamefont {Nahum},
  \citenamefont {Roy}, \citenamefont {Skinner},\ and\ \citenamefont
  {Ruhman}}]{Nahum2021AlltoAll}%
  \BibitemOpen
  \bibfield  {author} {\bibinfo {author} {\bibfnamefont {A.}~\bibnamefont
  {Nahum}}, \bibinfo {author} {\bibfnamefont {S.}~\bibnamefont {Roy}}, \bibinfo
  {author} {\bibfnamefont {B.}~\bibnamefont {Skinner}}, \ and\ \bibinfo
  {author} {\bibfnamefont {J.}~\bibnamefont {Ruhman}},\ }\href {\doibase
  10.1103/PRXQuantum.2.010352} {\bibfield  {journal} {\bibinfo  {journal} {PRX
  Quantum}\ }\textbf {\bibinfo {volume} {2}},\ \bibinfo {pages} {010352}
  (\bibinfo {year} {2021})}\BibitemShut {NoStop}%
\bibitem [{\citenamefont {{Bao}}\ \emph {et~al.}(2021)\citenamefont {{Bao}},
  \citenamefont {{Choi}},\ and\ \citenamefont
  {{Altman}}}]{BaoChoiEhud2021SymmetryEriched}%
  \BibitemOpen
  \bibfield  {author} {\bibinfo {author} {\bibfnamefont {Y.}~\bibnamefont
  {{Bao}}}, \bibinfo {author} {\bibfnamefont {S.}~\bibnamefont {{Choi}}}, \
  and\ \bibinfo {author} {\bibfnamefont {E.}~\bibnamefont {{Altman}}},\ }\href
  {\doibase 10.1016/j.aop.2021.168618} {\bibfield  {journal} {\bibinfo
  {journal} {Annals of Physics}\ }\textbf {\bibinfo {volume} {435}},\ \bibinfo
  {eid} {168618} (\bibinfo {year} {2021})},\ \Eprint
  {http://arxiv.org/abs/2102.09164} {arXiv:2102.09164 [cond-mat.stat-mech]}
  \BibitemShut {NoStop}%
\bibitem [{\citenamefont {{Turkeshi}}\ \emph {et~al.}(2020)\citenamefont
  {{Turkeshi}}, \citenamefont {{Fazio}},\ and\ \citenamefont
  {{Dalmonte}}}]{Turkeshi2020TwoD}%
  \BibitemOpen
  \bibfield  {author} {\bibinfo {author} {\bibfnamefont {X.}~\bibnamefont
  {{Turkeshi}}}, \bibinfo {author} {\bibfnamefont {R.}~\bibnamefont {{Fazio}}},
  \ and\ \bibinfo {author} {\bibfnamefont {M.}~\bibnamefont {{Dalmonte}}},\
  }\href {\doibase 10.1103/PhysRevB.102.014315} {\bibfield  {journal} {\bibinfo
   {journal} {\prb}\ }\textbf {\bibinfo {volume} {102}},\ \bibinfo {eid}
  {014315} (\bibinfo {year} {2020})},\ \Eprint
  {http://arxiv.org/abs/2007.02970} {arXiv:2007.02970 [cond-mat.stat-mech]}
  \BibitemShut {NoStop}%
\bibitem [{\citenamefont {{Buchhold}}\ \emph {et~al.}(2021)\citenamefont
  {{Buchhold}}, \citenamefont {{Minoguchi}}, \citenamefont {{Altland}},\ and\
  \citenamefont {{Diehl}}}]{Buchhold2021Dirac}%
  \BibitemOpen
  \bibfield  {author} {\bibinfo {author} {\bibfnamefont {M.}~\bibnamefont
  {{Buchhold}}}, \bibinfo {author} {\bibfnamefont {Y.}~\bibnamefont
  {{Minoguchi}}}, \bibinfo {author} {\bibfnamefont {A.}~\bibnamefont
  {{Altland}}}, \ and\ \bibinfo {author} {\bibfnamefont {S.}~\bibnamefont
  {{Diehl}}},\ }\href {\doibase 10.1103/PhysRevX.11.041004} {\bibfield
  {journal} {\bibinfo  {journal} {Physical Review X}\ }\textbf {\bibinfo
  {volume} {11}},\ \bibinfo {eid} {041004} (\bibinfo {year} {2021})},\ \Eprint
  {http://arxiv.org/abs/2102.08381} {arXiv:2102.08381 [cond-mat.stat-mech]}
  \BibitemShut {NoStop}%
\bibitem [{\citenamefont {{Barratt}}\ \emph {et~al.}(2022)\citenamefont
  {{Barratt}}, \citenamefont {{Agrawal}}, \citenamefont {{Gopalakrishnan}},
  \citenamefont {{Huse}}, \citenamefont {{Vasseur}},\ and\ \citenamefont
  {{Potter}}}]{Vasseur2022ChargeSharpen}%
  \BibitemOpen
  \bibfield  {author} {\bibinfo {author} {\bibfnamefont {F.}~\bibnamefont
  {{Barratt}}}, \bibinfo {author} {\bibfnamefont {U.}~\bibnamefont
  {{Agrawal}}}, \bibinfo {author} {\bibfnamefont {S.}~\bibnamefont
  {{Gopalakrishnan}}}, \bibinfo {author} {\bibfnamefont {D.~A.}\ \bibnamefont
  {{Huse}}}, \bibinfo {author} {\bibfnamefont {R.}~\bibnamefont {{Vasseur}}}, \
  and\ \bibinfo {author} {\bibfnamefont {A.~C.}\ \bibnamefont {{Potter}}},\
  }\href {\doibase 10.1103/PhysRevLett.129.120604} {\bibfield  {journal}
  {\bibinfo  {journal} {\prl}\ }\textbf {\bibinfo {volume} {129}},\ \bibinfo
  {eid} {120604} (\bibinfo {year} {2022})},\ \Eprint
  {http://arxiv.org/abs/2111.09336} {arXiv:2111.09336 [quant-ph]} \BibitemShut
  {NoStop}%
\bibitem [{\citenamefont {{Agrawal}}\ \emph {et~al.}(2022)\citenamefont
  {{Agrawal}}, \citenamefont {{Zabalo}}, \citenamefont {{Chen}}, \citenamefont
  {{Wilson}}, \citenamefont {{Potter}}, \citenamefont {{Pixley}}, \citenamefont
  {{Gopalakrishnan}},\ and\ \citenamefont {{Vasseur}}}]{Vasseur2022U1}%
  \BibitemOpen
  \bibfield  {author} {\bibinfo {author} {\bibfnamefont {U.}~\bibnamefont
  {{Agrawal}}}, \bibinfo {author} {\bibfnamefont {A.}~\bibnamefont {{Zabalo}}},
  \bibinfo {author} {\bibfnamefont {K.}~\bibnamefont {{Chen}}}, \bibinfo
  {author} {\bibfnamefont {J.~H.}\ \bibnamefont {{Wilson}}}, \bibinfo {author}
  {\bibfnamefont {A.~C.}\ \bibnamefont {{Potter}}}, \bibinfo {author}
  {\bibfnamefont {J.~H.}\ \bibnamefont {{Pixley}}}, \bibinfo {author}
  {\bibfnamefont {S.}~\bibnamefont {{Gopalakrishnan}}}, \ and\ \bibinfo
  {author} {\bibfnamefont {R.}~\bibnamefont {{Vasseur}}},\ }\href {\doibase
  10.1103/PhysRevX.12.041002} {\bibfield  {journal} {\bibinfo  {journal}
  {Physical Review X}\ }\textbf {\bibinfo {volume} {12}},\ \bibinfo {eid}
  {041002} (\bibinfo {year} {2022})},\ \Eprint
  {http://arxiv.org/abs/2107.10279} {arXiv:2107.10279 [cond-mat.dis-nn]}
  \BibitemShut {NoStop}%
\bibitem [{\citenamefont {Ippoliti}\ \emph {et~al.}(2022)\citenamefont
  {Ippoliti}, \citenamefont {Rakovszky},\ and\ \citenamefont
  {Khemani}}]{Ippoliti2022SpacetimeDual}%
  \BibitemOpen
  \bibfield  {author} {\bibinfo {author} {\bibfnamefont {M.}~\bibnamefont
  {Ippoliti}}, \bibinfo {author} {\bibfnamefont {T.}~\bibnamefont {Rakovszky}},
  \ and\ \bibinfo {author} {\bibfnamefont {V.}~\bibnamefont {Khemani}},\ }\href
  {\doibase 10.1103/PhysRevX.12.011045} {\bibfield  {journal} {\bibinfo
  {journal} {Phys. Rev. X}\ }\textbf {\bibinfo {volume} {12}},\ \bibinfo
  {pages} {011045} (\bibinfo {year} {2022})}\BibitemShut {NoStop}%
\bibitem [{\citenamefont {Zabalo}\ \emph {et~al.}(2022)\citenamefont {Zabalo},
  \citenamefont {Gullans}, \citenamefont {Wilson}, \citenamefont {Vasseur},
  \citenamefont {Ludwig}, \citenamefont {Gopalakrishnan}, \citenamefont
  {Huse},\ and\ \citenamefont {Pixley}}]{Pixley2022Multifractality}%
  \BibitemOpen
  \bibfield  {author} {\bibinfo {author} {\bibfnamefont {A.}~\bibnamefont
  {Zabalo}}, \bibinfo {author} {\bibfnamefont {M.~J.}\ \bibnamefont {Gullans}},
  \bibinfo {author} {\bibfnamefont {J.~H.}\ \bibnamefont {Wilson}}, \bibinfo
  {author} {\bibfnamefont {R.}~\bibnamefont {Vasseur}}, \bibinfo {author}
  {\bibfnamefont {A.~W.~W.}\ \bibnamefont {Ludwig}}, \bibinfo {author}
  {\bibfnamefont {S.}~\bibnamefont {Gopalakrishnan}}, \bibinfo {author}
  {\bibfnamefont {D.~A.}\ \bibnamefont {Huse}}, \ and\ \bibinfo {author}
  {\bibfnamefont {J.~H.}\ \bibnamefont {Pixley}},\ }\href {\doibase
  10.1103/PhysRevLett.128.050602} {\bibfield  {journal} {\bibinfo  {journal}
  {Phys. Rev. Lett.}\ }\textbf {\bibinfo {volume} {128}},\ \bibinfo {pages}
  {050602} (\bibinfo {year} {2022})}\BibitemShut {NoStop}%
\bibitem [{\citenamefont {{Li}}\ \emph {et~al.}(2021)\citenamefont {{Li}},
  \citenamefont {{Vasseur}}, \citenamefont {{Fisher}},\ and\ \citenamefont
  {{Ludwig}}}]{LiVasseurFisherLudwig2021}%
  \BibitemOpen
  \bibfield  {author} {\bibinfo {author} {\bibfnamefont {Y.}~\bibnamefont
  {{Li}}}, \bibinfo {author} {\bibfnamefont {R.}~\bibnamefont {{Vasseur}}},
  \bibinfo {author} {\bibfnamefont {M.~P.~A.}\ \bibnamefont {{Fisher}}}, \ and\
  \bibinfo {author} {\bibfnamefont {A.~W.~W.}\ \bibnamefont {{Ludwig}}},\
  }\href@noop {} {\bibfield  {journal} {\bibinfo  {journal} {arXiv e-prints}\
  ,\ \bibinfo {eid} {arXiv:2110.02988}} (\bibinfo {year} {2021})},\ \Eprint
  {http://arxiv.org/abs/2110.02988} {arXiv:2110.02988 [cond-mat.stat-mech]}
  \BibitemShut {NoStop}%
\bibitem [{\citenamefont {Jian}\ \emph {et~al.}(2022)\citenamefont {Jian},
  \citenamefont {Bauer}, \citenamefont {Keselman},\ and\ \citenamefont
  {Ludwig}}]{Jian2020}%
  \BibitemOpen
  \bibfield  {author} {\bibinfo {author} {\bibfnamefont {C.-M.}\ \bibnamefont
  {Jian}}, \bibinfo {author} {\bibfnamefont {B.}~\bibnamefont {Bauer}},
  \bibinfo {author} {\bibfnamefont {A.}~\bibnamefont {Keselman}}, \ and\
  \bibinfo {author} {\bibfnamefont {A.~W.~W.}\ \bibnamefont {Ludwig}},\ }\href
  {\doibase 10.1103/PhysRevB.106.134206} {\bibfield  {journal} {\bibinfo
  {journal} {Phys. Rev. B}\ }\textbf {\bibinfo {volume} {106}},\ \bibinfo
  {pages} {134206} (\bibinfo {year} {2022})}\BibitemShut {NoStop}%
\bibitem [{\citenamefont {{Cao}}\ \emph {et~al.}(2019)\citenamefont {{Cao}},
  \citenamefont {{Tilloy}},\ and\ \citenamefont {{De Luca}}}]{Cao2019FF}%
  \BibitemOpen
  \bibfield  {author} {\bibinfo {author} {\bibfnamefont {X.}~\bibnamefont
  {{Cao}}}, \bibinfo {author} {\bibfnamefont {A.}~\bibnamefont {{Tilloy}}}, \
  and\ \bibinfo {author} {\bibfnamefont {A.}~\bibnamefont {{De Luca}}},\ }\href
  {\doibase 10.21468/SciPostPhys.7.2.024} {\bibfield  {journal} {\bibinfo
  {journal} {SciPost Physics}\ }\textbf {\bibinfo {volume} {7}},\ \bibinfo
  {eid} {024} (\bibinfo {year} {2019})},\ \Eprint
  {http://arxiv.org/abs/1804.04638} {arXiv:1804.04638 [cond-mat.stat-mech]}
  \BibitemShut {NoStop}%
\bibitem [{\citenamefont {{Nahum}}\ and\ \citenamefont
  {{Skinner}}(2020)}]{NahumSkinnerMajDefect2020FF}%
  \BibitemOpen
  \bibfield  {author} {\bibinfo {author} {\bibfnamefont {A.}~\bibnamefont
  {{Nahum}}}\ and\ \bibinfo {author} {\bibfnamefont {B.}~\bibnamefont
  {{Skinner}}},\ }\href {\doibase 10.1103/PhysRevResearch.2.023288} {\bibfield
  {journal} {\bibinfo  {journal} {Physical Review Research}\ }\textbf {\bibinfo
  {volume} {2}},\ \bibinfo {eid} {023288} (\bibinfo {year} {2020})},\ \Eprint
  {http://arxiv.org/abs/1911.11169} {arXiv:1911.11169 [cond-mat.stat-mech]}
  \BibitemShut {NoStop}%
\bibitem [{\citenamefont {{Chen}}\ \emph {et~al.}(2020)\citenamefont {{Chen}},
  \citenamefont {{Li}}, \citenamefont {{Fisher}},\ and\ \citenamefont
  {{Lucas}}}]{ChenLiFisherLucas2020FF}%
  \BibitemOpen
  \bibfield  {author} {\bibinfo {author} {\bibfnamefont {X.}~\bibnamefont
  {{Chen}}}, \bibinfo {author} {\bibfnamefont {Y.}~\bibnamefont {{Li}}},
  \bibinfo {author} {\bibfnamefont {M.~P.~A.}\ \bibnamefont {{Fisher}}}, \ and\
  \bibinfo {author} {\bibfnamefont {A.}~\bibnamefont {{Lucas}}},\ }\href
  {\doibase 10.1103/PhysRevResearch.2.033017} {\bibfield  {journal} {\bibinfo
  {journal} {Physical Review Research}\ }\textbf {\bibinfo {volume} {2}},\
  \bibinfo {eid} {033017} (\bibinfo {year} {2020})},\ \Eprint
  {http://arxiv.org/abs/2004.09577} {arXiv:2004.09577 [quant-ph]} \BibitemShut
  {NoStop}%
\bibitem [{\citenamefont {Lu}\ and\ \citenamefont
  {Grover}(2021)}]{LuGrover2021SpacetimeDualFF}%
  \BibitemOpen
  \bibfield  {author} {\bibinfo {author} {\bibfnamefont {T.-C.}\ \bibnamefont
  {Lu}}\ and\ \bibinfo {author} {\bibfnamefont {T.}~\bibnamefont {Grover}},\
  }\href {\doibase 10.1103/PRXQuantum.2.040319} {\bibfield  {journal} {\bibinfo
   {journal} {PRX Quantum}\ }\textbf {\bibinfo {volume} {2}},\ \bibinfo {pages}
  {040319} (\bibinfo {year} {2021})},\ \Eprint
  {http://arxiv.org/abs/2103.06356} {arXiv:2103.06356 [quant-ph]} \BibitemShut
  {NoStop}%
\bibitem [{\citenamefont {{M{\"u}ller}}\ \emph {et~al.}(2022)\citenamefont
  {{M{\"u}ller}}, \citenamefont {{Diehl}},\ and\ \citenamefont
  {{Buchhold}}}]{Buchhold2022FF}%
  \BibitemOpen
  \bibfield  {author} {\bibinfo {author} {\bibfnamefont {T.}~\bibnamefont
  {{M{\"u}ller}}}, \bibinfo {author} {\bibfnamefont {S.}~\bibnamefont
  {{Diehl}}}, \ and\ \bibinfo {author} {\bibfnamefont {M.}~\bibnamefont
  {{Buchhold}}},\ }\href {\doibase 10.1103/PhysRevLett.128.010605} {\bibfield
  {journal} {\bibinfo  {journal} {\prl}\ }\textbf {\bibinfo {volume} {128}},\
  \bibinfo {eid} {010605} (\bibinfo {year} {2022})},\ \Eprint
  {http://arxiv.org/abs/2105.08076} {arXiv:2105.08076 [cond-mat.stat-mech]}
  \BibitemShut {NoStop}%
\bibitem [{\citenamefont {Tang}\ \emph {et~al.}(2021)\citenamefont {Tang},
  \citenamefont {Chen},\ and\ \citenamefont {Zhu}}]{TangChenZhu2021FF}%
  \BibitemOpen
  \bibfield  {author} {\bibinfo {author} {\bibfnamefont {Q.}~\bibnamefont
  {Tang}}, \bibinfo {author} {\bibfnamefont {X.}~\bibnamefont {Chen}}, \ and\
  \bibinfo {author} {\bibfnamefont {W.}~\bibnamefont {Zhu}},\ }\href {\doibase
  10.1103/PhysRevB.103.174303} {\bibfield  {journal} {\bibinfo  {journal}
  {Phys. Rev. B}\ }\textbf {\bibinfo {volume} {103}},\ \bibinfo {pages}
  {174303} (\bibinfo {year} {2021})}\BibitemShut {NoStop}%
\bibitem [{\citenamefont {{Lavasani}}\ \emph {et~al.}(2022)\citenamefont
  {{Lavasani}}, \citenamefont {{Luo}},\ and\ \citenamefont
  {{Vijay}}}]{LavasaniLuoVijay2022}%
  \BibitemOpen
  \bibfield  {author} {\bibinfo {author} {\bibfnamefont {A.}~\bibnamefont
  {{Lavasani}}}, \bibinfo {author} {\bibfnamefont {Z.-X.}\ \bibnamefont
  {{Luo}}}, \ and\ \bibinfo {author} {\bibfnamefont {S.}~\bibnamefont
  {{Vijay}}},\ }\href@noop {} {\bibfield  {journal} {\bibinfo  {journal} {arXiv
  e-prints}\ ,\ \bibinfo {eid} {arXiv:2207.02877}} (\bibinfo {year} {2022})},\
  \Eprint {http://arxiv.org/abs/2207.02877} {arXiv:2207.02877
  [cond-mat.str-el]} \BibitemShut {NoStop}%
\bibitem [{\citenamefont {{Sriram}}\ \emph {et~al.}(2022)\citenamefont
  {{Sriram}}, \citenamefont {{Rakovszky}}, \citenamefont {{Khemani}},\ and\
  \citenamefont {{Ippoliti}}}]{Sriram2022KitaevModel}%
  \BibitemOpen
  \bibfield  {author} {\bibinfo {author} {\bibfnamefont {A.}~\bibnamefont
  {{Sriram}}}, \bibinfo {author} {\bibfnamefont {T.}~\bibnamefont
  {{Rakovszky}}}, \bibinfo {author} {\bibfnamefont {V.}~\bibnamefont
  {{Khemani}}}, \ and\ \bibinfo {author} {\bibfnamefont {M.}~\bibnamefont
  {{Ippoliti}}},\ }\href@noop {} {\bibfield  {journal} {\bibinfo  {journal}
  {arXiv e-prints}\ ,\ \bibinfo {eid} {arXiv:2207.07096}} (\bibinfo {year}
  {2022})},\ \Eprint {http://arxiv.org/abs/2207.07096} {arXiv:2207.07096
  [quant-ph]} \BibitemShut {NoStop}%
\bibitem [{\citenamefont {{Potter}}\ and\ \citenamefont
  {{Vasseur}}(2021)}]{PotterVasseur2021EntanglementDynamics}%
  \BibitemOpen
  \bibfield  {author} {\bibinfo {author} {\bibfnamefont {A.~C.}\ \bibnamefont
  {{Potter}}}\ and\ \bibinfo {author} {\bibfnamefont {R.}~\bibnamefont
  {{Vasseur}}},\ }\href@noop {} {\bibfield  {journal} {\bibinfo  {journal}
  {arXiv e-prints}\ ,\ \bibinfo {eid} {arXiv:2111.08018}} (\bibinfo {year}
  {2021})},\ \Eprint {http://arxiv.org/abs/2111.08018} {arXiv:2111.08018
  [quant-ph]} \BibitemShut {NoStop}%
\bibitem [{\citenamefont {{Fisher}}\ \emph {et~al.}(2022)\citenamefont
  {{Fisher}}, \citenamefont {{Khemani}}, \citenamefont {{Nahum}},\ and\
  \citenamefont {{Vijay}}}]{FisherKhemaniNahumVijay2022RQC}%
  \BibitemOpen
  \bibfield  {author} {\bibinfo {author} {\bibfnamefont {M.~P.~A.}\
  \bibnamefont {{Fisher}}}, \bibinfo {author} {\bibfnamefont {V.}~\bibnamefont
  {{Khemani}}}, \bibinfo {author} {\bibfnamefont {A.}~\bibnamefont {{Nahum}}},
  \ and\ \bibinfo {author} {\bibfnamefont {S.}~\bibnamefont {{Vijay}}},\
  }\href@noop {} {\bibfield  {journal} {\bibinfo  {journal} {arXiv e-prints}\
  ,\ \bibinfo {eid} {arXiv:2207.14280}} (\bibinfo {year} {2022})},\ \Eprint
  {http://arxiv.org/abs/2207.14280} {arXiv:2207.14280 [quant-ph]} \BibitemShut
  {NoStop}%
\bibitem [{\citenamefont {{Alberton}}\ \emph {et~al.}(2021)\citenamefont
  {{Alberton}}, \citenamefont {{Buchhold}},\ and\ \citenamefont
  {{Diehl}}}]{DiehlPRL2021FF}%
  \BibitemOpen
  \bibfield  {author} {\bibinfo {author} {\bibfnamefont {O.}~\bibnamefont
  {{Alberton}}}, \bibinfo {author} {\bibfnamefont {M.}~\bibnamefont
  {{Buchhold}}}, \ and\ \bibinfo {author} {\bibfnamefont {S.}~\bibnamefont
  {{Diehl}}},\ }\href {\doibase 10.1103/PhysRevLett.126.170602} {\bibfield
  {journal} {\bibinfo  {journal} {\prl}\ }\textbf {\bibinfo {volume} {126}},\
  \bibinfo {eid} {170602} (\bibinfo {year} {2021})},\ \Eprint
  {http://arxiv.org/abs/2005.09722} {arXiv:2005.09722 [cond-mat.stat-mech]}
  \BibitemShut {NoStop}%
\bibitem [{\citenamefont {Altland}\ and\ \citenamefont
  {Zirnbauer}(1997)}]{AltlandZirnbauerPRB1997}%
  \BibitemOpen
  \bibfield  {author} {\bibinfo {author} {\bibfnamefont {A.}~\bibnamefont
  {Altland}}\ and\ \bibinfo {author} {\bibfnamefont {M.~R.}\ \bibnamefont
  {Zirnbauer}},\ }\href {\doibase 10.1103/PhysRevB.55.1142} {\bibfield
  {journal} {\bibinfo  {journal} {Phys. Rev. B}\ }\textbf {\bibinfo {volume}
  {55}},\ \bibinfo {pages} {1142} (\bibinfo {year} {1997})}\BibitemShut
  {NoStop}%
\bibitem [{\citenamefont {Ryu}\ \emph {et~al.}(2010)\citenamefont {Ryu},
  \citenamefont {Schnyder}, \citenamefont {Furusaki},\ and\ \citenamefont
  {Ludwig}}]{RyuSchnyderFurusakiLudwigNJPhys2010}%
  \BibitemOpen
  \bibfield  {author} {\bibinfo {author} {\bibfnamefont {S.}~\bibnamefont
  {Ryu}}, \bibinfo {author} {\bibfnamefont {A.~P.}\ \bibnamefont {Schnyder}},
  \bibinfo {author} {\bibfnamefont {A.}~\bibnamefont {Furusaki}}, \ and\
  \bibinfo {author} {\bibfnamefont {A.~W.~W.}\ \bibnamefont {Ludwig}},\ }\href
  {\doibase 10.1088/1367-2630/12/6/065010} {\bibfield  {journal} {\bibinfo
  {journal} {New Journal of Physics}\ }\textbf {\bibinfo {volume} {12}},\
  \bibinfo {pages} {065010} (\bibinfo {year} {2010})}\BibitemShut {NoStop}%
\bibitem [{Note1()}]{Note1}%
  \BibitemOpen
  \bibinfo {note} {As an aside, we note that the term ``forced measurement''
  has been used with a different meaning in the literature of measurement-based
  topological quantum computation, where a measurement outcome is forced using
  a ``repeat-until-success'' approach implemented by a
  probabilistically-determined adaptive series of operations. ~\cite
  {BondersonFreedmanNayakMeasurementOnlyPRL2008,BondersonFreedmanNayakMeasurementOnlyAnnPhys2009,TranBocharovBauerBondersonSciPost2020}}\BibitemShut
  {NoStop}%
\bibitem [{\citenamefont {Vasseur}\ \emph {et~al.}(2019)\citenamefont
  {Vasseur}, \citenamefont {Potter}, \citenamefont {You},\ and\ \citenamefont
  {Ludwig}}]{VasseurPotterYouLudwigRTNPRB2019}%
  \BibitemOpen
  \bibfield  {author} {\bibinfo {author} {\bibfnamefont {R.}~\bibnamefont
  {Vasseur}}, \bibinfo {author} {\bibfnamefont {A.~C.}\ \bibnamefont {Potter}},
  \bibinfo {author} {\bibfnamefont {Y.-Z.}\ \bibnamefont {You}}, \ and\
  \bibinfo {author} {\bibfnamefont {A.~W.~W.}\ \bibnamefont {Ludwig}},\ }\href
  {\doibase 10.1103/PhysRevB.100.134203} {\bibfield  {journal} {\bibinfo
  {journal} {Phys. Rev. B}\ }\textbf {\bibinfo {volume} {100}},\ \bibinfo
  {pages} {134203} (\bibinfo {year} {2019})}\BibitemShut {NoStop}%
\bibitem [{\citenamefont {{Ludwig}}(1990)}]{LudwigHierarchies1990}%
  \BibitemOpen
  \bibfield  {author} {\bibinfo {author} {\bibfnamefont {A.~W.~W.}\
  \bibnamefont {{Ludwig}}},\ }\href {\doibase
  https://doi.org/10.1016/0550-3213(90)90126-X} {\bibfield  {journal} {\bibinfo
   {journal} {Nuclear Physics B}\ }\textbf {\bibinfo {volume} {330}},\ \bibinfo
  {pages} {639} (\bibinfo {year} {1990})}\BibitemShut {NoStop}%
\bibitem [{\citenamefont {{Fu}}\ and\ \citenamefont
  {{Kane}}(2012)}]{FuKane2012Sympletic}%
  \BibitemOpen
  \bibfield  {author} {\bibinfo {author} {\bibfnamefont {L.}~\bibnamefont
  {{Fu}}}\ and\ \bibinfo {author} {\bibfnamefont {C.~L.}\ \bibnamefont
  {{Kane}}},\ }\href {\doibase 10.1103/PhysRevLett.109.246605} {\bibfield
  {journal} {\bibinfo  {journal} {\prl}\ }\textbf {\bibinfo {volume} {109}},\
  \bibinfo {eid} {246605} (\bibinfo {year} {2012})},\ \Eprint
  {http://arxiv.org/abs/1208.3442} {arXiv:1208.3442 [cond-mat.mes-hall]}
  \BibitemShut {NoStop}%
\bibitem [{\citenamefont {{Bravyi}}(2004)}]{Bravyi2004FermionicOptics}%
  \BibitemOpen
  \bibfield  {author} {\bibinfo {author} {\bibfnamefont {S.}~\bibnamefont
  {{Bravyi}}},\ }\href@noop {} {\bibfield  {journal} {\bibinfo  {journal}
  {arXiv e-prints}\ ,\ \bibinfo {eid} {quant-ph/0404180}} (\bibinfo {year}
  {2004})},\ \Eprint {http://arxiv.org/abs/quant-ph/0404180}
  {arXiv:quant-ph/0404180 [quant-ph]} \BibitemShut {NoStop}%
\bibitem [{Note2()}]{Note2}%
  \BibitemOpen
  \bibinfo {note} {The more detailed statement is that any generalized
  measurements can be implemented by combining unitary operations and
  projective measurements (possibly with the help of ancillary degrees of
  freedom). See for example Ref. \protect \rev@citealp {watrous2018theory} (in
  particular Section 2.3 and Theorem 2.42).}\BibitemShut {Stop}%
\bibitem [{Note3()}]{Note3}%
  \BibitemOpen
  \bibinfo {note} {The symmetries of the AZ classification refer to that of the
  system with a static Hamiltonian in a spatial dimension equal to the
  space-time dimension of the circuit, which appears in the aforementioned
  correspondence. Due to the correspondence, the classification of the systems
  with static Hamiltonian is inherited by the circuits, but the symmetry is
  generally not. The monitored Gaussian circuits in this paper respect no
  symmetry other than fermion parity.}\BibitemShut {Stop}%
\bibitem [{\citenamefont {{Li}}\ \emph {et~al.}(2020)\citenamefont {{Li}},
  \citenamefont {{Chen}}, \citenamefont {{Ludwig}},\ and\ \citenamefont
  {{Fisher}}}]{Li2020ConformalSym}%
  \BibitemOpen
  \bibfield  {author} {\bibinfo {author} {\bibfnamefont {Y.}~\bibnamefont
  {{Li}}}, \bibinfo {author} {\bibfnamefont {X.}~\bibnamefont {{Chen}}},
  \bibinfo {author} {\bibfnamefont {A.~W.~W.}\ \bibnamefont {{Ludwig}}}, \ and\
  \bibinfo {author} {\bibfnamefont {M.~P.~A.}\ \bibnamefont {{Fisher}}},\
  }\href@noop {} {\bibfield  {journal} {\bibinfo  {journal} {arXiv e-prints}\
  ,\ \bibinfo {eid} {arXiv:2003.12721}} (\bibinfo {year} {2020})},\ \Eprint
  {http://arxiv.org/abs/2003.12721} {arXiv:2003.12721 [quant-ph]} \BibitemShut
  {NoStop}%
\bibitem [{Note4()}]{Note4}%
  \BibitemOpen
  \bibinfo {note} {Symmetry class DIII describes the Bogoliubov-deGennes
  Hamiltonian of a time-reversal invariant superconductor (or insulator with
  charge conservation).}\BibitemShut {Stop}%
\bibitem [{\citenamefont {{Fulga}}\ \emph {et~al.}(2012)\citenamefont
  {{Fulga}}, \citenamefont {{Akhmerov}}, \citenamefont {{Tworzyd{\l}o}},
  \citenamefont {{B{\'e}ri}},\ and\ \citenamefont
  {{Beenakker}}}]{Fulga2012DIIIMIT}%
  \BibitemOpen
  \bibfield  {author} {\bibinfo {author} {\bibfnamefont {I.~C.}\ \bibnamefont
  {{Fulga}}}, \bibinfo {author} {\bibfnamefont {A.~R.}\ \bibnamefont
  {{Akhmerov}}}, \bibinfo {author} {\bibfnamefont {J.}~\bibnamefont
  {{Tworzyd{\l}o}}}, \bibinfo {author} {\bibfnamefont {B.}~\bibnamefont
  {{B{\'e}ri}}}, \ and\ \bibinfo {author} {\bibfnamefont {C.~W.~J.}\
  \bibnamefont {{Beenakker}}},\ }\href {\doibase 10.1103/PhysRevB.86.054505}
  {\bibfield  {journal} {\bibinfo  {journal} {\prb}\ }\textbf {\bibinfo
  {volume} {86}},\ \bibinfo {eid} {054505} (\bibinfo {year} {2012})},\ \Eprint
  {http://arxiv.org/abs/1205.1441} {arXiv:1205.1441 [cond-mat.mes-hall]}
  \BibitemShut {NoStop}%
\bibitem [{\citenamefont {Schnyder}\ \emph {et~al.}(2008)\citenamefont
  {Schnyder}, \citenamefont {Ryu}, \citenamefont {Furusaki},\ and\
  \citenamefont
  {Ludwig}}]{SchnyderRyuFurusakiLudwigClassificationPhysRevB2008}%
  \BibitemOpen
  \bibfield  {author} {\bibinfo {author} {\bibfnamefont {A.~P.}\ \bibnamefont
  {Schnyder}}, \bibinfo {author} {\bibfnamefont {S.}~\bibnamefont {Ryu}},
  \bibinfo {author} {\bibfnamefont {A.}~\bibnamefont {Furusaki}}, \ and\
  \bibinfo {author} {\bibfnamefont {A.~W.~W.}\ \bibnamefont {Ludwig}},\ }\href
  {\doibase 10.1103/PhysRevB.78.195125} {\bibfield  {journal} {\bibinfo
  {journal} {Phys. Rev. B}\ }\textbf {\bibinfo {volume} {78}},\ \bibinfo
  {pages} {195125} (\bibinfo {year} {2008})}\BibitemShut {NoStop}%
\bibitem [{Note5()}]{Note5}%
  \BibitemOpen
  \bibinfo {note} {The mechanism for the appearance of the critical phase in
  symmetry class DIII, as explained in Ref.~\protect \rev@citealp
  {Fulga2012DIIIMIT}, following Ref.~\protect \rev@citealp
  {ChalkerEtAlThermalMetalPRB65} for class D, results in the fact that in this
  phase the field variable $O({\protect \vec r})$ resides in the component of
  the orthogonal group ${\protect \rm O}(n)$ which is connected to the
  identity, i.e. in ${\protect \rm SO}(n)$. Therefore, the critical phase is
  described by the ${\protect \rm SO}(n)$ and not by the ${\protect \rm O}(n)$
  principal chiral NLSM. As discussed below, the transition out of this
  critical phase into the adjacent area-law phases is driven by proliferation
  of topological defects in the ${\protect \rm SO}(n)$ principal chiral
  NLSM}\BibitemShut {NoStop}%
\bibitem [{Note6()}]{Note6}%
  \BibitemOpen
  \bibinfo {note} {Owning to the $n \to 0$ limit (or a corresponding
  supersymmetric formulation, briefly alluded to below), the Mermin-Wagner
  theorem is known not to apply.}\BibitemShut {Stop}%
\bibitem [{Note7()}]{Note7}%
  \BibitemOpen
  \bibinfo {note} {E.g., compare for the system discussed in the present paper,
  Fig. 3 of Ref. \protect \rev@citealp {Fulga2012DIIIMIT}, with the phase
  diagram in Fig. 11 of the symplectic class in Ref. \protect \rev@citealp
  {ObuseFurusakiETAL2007} (with tuning parameter $x$).}\BibitemShut {Stop}%
\bibitem [{\citenamefont {Polyakov}(1987)}]{PolyakovBook1987}%
  \BibitemOpen
  \bibfield  {author} {\bibinfo {author} {\bibfnamefont {A.}~\bibnamefont
  {Polyakov}},\ }\href {https://books.google.com/books?id=3VcIvgEACAAJ} {\emph
  {\bibinfo {title} {Gauge Fields and Strings}}},\ Contemporary concepts in
  physics\ (\bibinfo  {publisher} {Harwood Academic Publishers},\ \bibinfo
  {year} {1987})\BibitemShut {NoStop}%
\bibitem [{Note8()}]{Note8}%
  \BibitemOpen
  \bibinfo {note} {The coupling constant $g$ of the NLSM is marginally
  irrelevant in both limits $n\rightarrow 0$ and $n \rightarrow
  1$.}\BibitemShut {Stop}%
\bibitem [{Note9()}]{Note9}%
  \BibitemOpen
  \bibinfo {note} {We note in passing that, as is well known, the replica
  limits $n\to 0$ and $n\to 1$ can be avoided by using a different formulation
  in terms of supersymmetry[Ref(Efetov)].}\BibitemShut {Stop}%
\bibitem [{Note10()}]{Note10}%
  \BibitemOpen
  \bibinfo {note} {There is also a simpler loop model 'without crossings',
  where the ``swap gates'' (ii) are absent. This model has no critical phase,
  but instead an isolated transition between two area law phases which is in
  the well-studied universality class of two-dimensional percolation. Its
  universality class is manifestly different from the transitions discussed in
  the present paper, and we will not come back to this simpler loop model in
  this paper.}\BibitemShut {Stop}%
\bibitem [{\citenamefont {Nahum}\ \emph {et~al.}(2013)\citenamefont {Nahum},
  \citenamefont {Serna}, \citenamefont {Somoza},\ and\ \citenamefont
  {Ortu\~no}}]{Nahum2013LoopWithCrossing}%
  \BibitemOpen
  \bibfield  {author} {\bibinfo {author} {\bibfnamefont {A.}~\bibnamefont
  {Nahum}}, \bibinfo {author} {\bibfnamefont {P.}~\bibnamefont {Serna}},
  \bibinfo {author} {\bibfnamefont {A.~M.}\ \bibnamefont {Somoza}}, \ and\
  \bibinfo {author} {\bibfnamefont {M.}~\bibnamefont {Ortu\~no}},\ }\href
  {\doibase 10.1103/PhysRevB.87.184204} {\bibfield  {journal} {\bibinfo
  {journal} {Phys. Rev. B}\ }\textbf {\bibinfo {volume} {87}},\ \bibinfo
  {pages} {184204} (\bibinfo {year} {2013})}\BibitemShut {NoStop}%
\bibitem [{\citenamefont {Nahum}\ and\ \citenamefont
  {Chalker}(2012)}]{NahumChalkerVortexLinesPRE85}%
  \BibitemOpen
  \bibfield  {author} {\bibinfo {author} {\bibfnamefont {A.}~\bibnamefont
  {Nahum}}\ and\ \bibinfo {author} {\bibfnamefont {J.~T.}\ \bibnamefont
  {Chalker}},\ }\href {\doibase 10.1103/PhysRevE.85.031141} {\bibfield
  {journal} {\bibinfo  {journal} {Phys. Rev. E}\ }\textbf {\bibinfo {volume}
  {85}},\ \bibinfo {pages} {031141} (\bibinfo {year} {2012})}\BibitemShut
  {NoStop}%
\bibitem [{\citenamefont {Efetov}(1996)}]{BookEfetov_1996}%
  \BibitemOpen
  \bibfield  {author} {\bibinfo {author} {\bibfnamefont {K.}~\bibnamefont
  {Efetov}},\ }\href {\doibase 10.1017/CBO9780511573057} {\emph {\bibinfo
  {title} {Supersymmetry in Disorder and Chaos}}}\ (\bibinfo  {publisher}
  {Cambridge University Press},\ \bibinfo {year} {1996})\BibitemShut {NoStop}%
\bibitem [{Note11()}]{Note11}%
  \BibitemOpen
  \bibinfo {note} {The latter property represents, as briefly summarized in
  Appendix~\ref {LabelSectionAppendixMultifractal}, a wealth of universal data.
  These can also be encoded in a scaling form of the probability distribution
  characterized by a universal function~\cite
  {LudwigHierarchies1990,Pixley2022Multifractality}.}\BibitemShut {Stop}%
\bibitem [{Note12()}]{Note12}%
  \BibitemOpen
  \bibinfo {note} {The sole fact that one model contains only critical
  exponents describing the average (or a few low-order moments) of an
  observable while another model describes rich scaling behavior featuring
  unequal average and typical exponents of the same observable, does not in
  itself imply that the transitions in the two models are in unrelated
  universality classes. However, while two such models could lie in the same
  universality class if the former model forms a subsector of the latter model
  containing just a subset of the critical exponents of the latter model (as it
  is the case in the model discussed in Ref.~\protect \rev@citealp
  {GruzbergLudwigReadPRL1999}), such a situation is {\protect \it not} realized
  in the case discussed in the present paper because the NLSM for the
  loop-model with crossings is invariant under a different symmetry than the
  NLSM for the generic monitored Gaussian circuit discussed in this paper, as
  detailed above in this Section.}\BibitemShut {Stop}%
\bibitem [{\citenamefont {Shapourian}\ \emph {et~al.}(2022)\citenamefont
  {Shapourian}, \citenamefont {Jian}, \citenamefont {Bauer},\ and\
  \citenamefont {Ludwig}}]{HassanAPSabstract}%
  \BibitemOpen
  \bibfield  {author} {\bibinfo {author} {\bibfnamefont {H.}~\bibnamefont
  {Shapourian}}, \bibinfo {author} {\bibfnamefont {C.-M.}\ \bibnamefont
  {Jian}}, \bibinfo {author} {\bibfnamefont {B.}~\bibnamefont {Bauer}}, \ and\
  \bibinfo {author} {\bibfnamefont {A.~W.~W.}\ \bibnamefont {Ludwig}},\ }\href
  {https://meetings.aps.org/Meeting/MAR22/Session/S50.1} {\bibfield  {journal}
  {\bibinfo  {journal} {Bulletin of the American Physical Society: March
  Meeting 2022, Abstract: S50.00001}\ } (\bibinfo {year} {2022})}\BibitemShut
  {NoStop}%
\bibitem [{\citenamefont {{Merritt}}\ and\ \citenamefont
  {{Fidkowski}}(2022)}]{Lukasz2022}%
  \BibitemOpen
  \bibfield  {author} {\bibinfo {author} {\bibfnamefont {J.}~\bibnamefont
  {{Merritt}}}\ and\ \bibinfo {author} {\bibfnamefont {L.}~\bibnamefont
  {{Fidkowski}}},\ }\href@noop {} {\bibfield  {journal} {\bibinfo  {journal}
  {arXiv e-prints}\ ,\ \bibinfo {eid} {arXiv:2210.05681}} (\bibinfo {year}
  {2022})},\ \Eprint {http://arxiv.org/abs/2210.05681} {arXiv:2210.05681
  [cond-mat.str-el]} \BibitemShut {NoStop}%
\bibitem [{\citenamefont {Duplantier}\ and\ \citenamefont
  {Ludwig}(1991)}]{DuplantierLudwigPRL1991}%
  \BibitemOpen
  \bibfield  {author} {\bibinfo {author} {\bibfnamefont {B.}~\bibnamefont
  {Duplantier}}\ and\ \bibinfo {author} {\bibfnamefont {A.~W.~W.}\ \bibnamefont
  {Ludwig}},\ }\href {\doibase 10.1103/PhysRevLett.66.247} {\bibfield
  {journal} {\bibinfo  {journal} {Phys. Rev. Lett.}\ }\textbf {\bibinfo
  {volume} {66}},\ \bibinfo {pages} {247} (\bibinfo {year} {1991})}\BibitemShut
  {NoStop}%
\bibitem [{Note13()}]{Note13}%
  \BibitemOpen
  \bibinfo {note} {When all moments are identical, one may not want to refer to
  this as ``multifractal'', as this is in a way a ``trivial'' case where there
  is really only one moment, namely the 1st moment.}\BibitemShut {Stop}%
\bibitem [{\citenamefont {Bonderson}\ \emph {et~al.}(2008)\citenamefont
  {Bonderson}, \citenamefont {Freedman},\ and\ \citenamefont
  {Nayak}}]{BondersonFreedmanNayakMeasurementOnlyPRL2008}%
  \BibitemOpen
  \bibfield  {author} {\bibinfo {author} {\bibfnamefont {P.}~\bibnamefont
  {Bonderson}}, \bibinfo {author} {\bibfnamefont {M.}~\bibnamefont {Freedman}},
  \ and\ \bibinfo {author} {\bibfnamefont {C.}~\bibnamefont {Nayak}},\ }\href
  {\doibase 10.1103/PhysRevLett.101.010501} {\bibfield  {journal} {\bibinfo
  {journal} {Phys. Rev. Lett.}\ }\textbf {\bibinfo {volume} {101}},\ \bibinfo
  {pages} {010501} (\bibinfo {year} {2008})}\BibitemShut {NoStop}%
\bibitem [{\citenamefont {Bonderson}\ \emph {et~al.}(2009)\citenamefont
  {Bonderson}, \citenamefont {Freedman},\ and\ \citenamefont
  {Nayak}}]{BondersonFreedmanNayakMeasurementOnlyAnnPhys2009}%
  \BibitemOpen
  \bibfield  {author} {\bibinfo {author} {\bibfnamefont {P.}~\bibnamefont
  {Bonderson}}, \bibinfo {author} {\bibfnamefont {M.}~\bibnamefont {Freedman}},
  \ and\ \bibinfo {author} {\bibfnamefont {C.}~\bibnamefont {Nayak}},\ }\href
  {\doibase https://doi.org/10.1016/j.aop.2008.09.009} {\bibfield  {journal}
  {\bibinfo  {journal} {Annals of Physics}\ }\textbf {\bibinfo {volume}
  {324}},\ \bibinfo {pages} {787} (\bibinfo {year} {2009})}\BibitemShut
  {NoStop}%
\bibitem [{\citenamefont {Tran}\ \emph {et~al.}(2020)\citenamefont {Tran},
  \citenamefont {Bocharov}, \citenamefont {Bauer},\ and\ \citenamefont
  {Bonderson}}]{TranBocharovBauerBondersonSciPost2020}%
  \BibitemOpen
  \bibfield  {author} {\bibinfo {author} {\bibfnamefont {A.}~\bibnamefont
  {Tran}}, \bibinfo {author} {\bibfnamefont {A.}~\bibnamefont {Bocharov}},
  \bibinfo {author} {\bibfnamefont {B.}~\bibnamefont {Bauer}}, \ and\ \bibinfo
  {author} {\bibfnamefont {P.}~\bibnamefont {Bonderson}},\ }\href {\doibase
  10.21468/SciPostPhys.8.6.091} {\bibfield  {journal} {\bibinfo  {journal}
  {SciPost Phys.}\ }\textbf {\bibinfo {volume} {8}},\ \bibinfo {pages} {091}
  (\bibinfo {year} {2020})}\BibitemShut {NoStop}%
\bibitem [{\citenamefont {Watrous}(2018)}]{watrous2018theory}%
  \BibitemOpen
  \bibfield  {author} {\bibinfo {author} {\bibfnamefont {J.}~\bibnamefont
  {Watrous}},\ }\href@noop {} {\emph {\bibinfo {title} {The theory of quantum
  information}}}\ (\bibinfo  {publisher} {Cambridge University Press},\
  \bibinfo {year} {2018})\BibitemShut {NoStop}%
\bibitem [{\citenamefont {Chalker}\ \emph {et~al.}(2001)\citenamefont
  {Chalker}, \citenamefont {Read}, \citenamefont {Kagalovsky}, \citenamefont
  {Horovitz}, \citenamefont {Avishai},\ and\ \citenamefont
  {Ludwig}}]{ChalkerEtAlThermalMetalPRB65}%
  \BibitemOpen
  \bibfield  {author} {\bibinfo {author} {\bibfnamefont {J.~T.}\ \bibnamefont
  {Chalker}}, \bibinfo {author} {\bibfnamefont {N.}~\bibnamefont {Read}},
  \bibinfo {author} {\bibfnamefont {V.}~\bibnamefont {Kagalovsky}}, \bibinfo
  {author} {\bibfnamefont {B.}~\bibnamefont {Horovitz}}, \bibinfo {author}
  {\bibfnamefont {Y.}~\bibnamefont {Avishai}}, \ and\ \bibinfo {author}
  {\bibfnamefont {A.~W.~W.}\ \bibnamefont {Ludwig}},\ }\href {\doibase
  10.1103/PhysRevB.65.012506} {\bibfield  {journal} {\bibinfo  {journal} {Phys.
  Rev. B}\ }\textbf {\bibinfo {volume} {65}},\ \bibinfo {pages} {012506}
  (\bibinfo {year} {2001})}\BibitemShut {NoStop}%
\bibitem [{\citenamefont {{Obuse}}\ \emph {et~al.}(2007)\citenamefont
  {{Obuse}}, \citenamefont {{Furusaki}}, \citenamefont {{Ryu}},\ and\
  \citenamefont {{Mudry}}}]{ObuseFurusakiETAL2007}%
  \BibitemOpen
  \bibfield  {author} {\bibinfo {author} {\bibfnamefont {H.}~\bibnamefont
  {{Obuse}}}, \bibinfo {author} {\bibfnamefont {A.}~\bibnamefont {{Furusaki}}},
  \bibinfo {author} {\bibfnamefont {S.}~\bibnamefont {{Ryu}}}, \ and\ \bibinfo
  {author} {\bibfnamefont {C.}~\bibnamefont {{Mudry}}},\ }\href {\doibase
  10.1103/PhysRevB.76.075301} {\bibfield  {journal} {\bibinfo  {journal}
  {\prb}\ }\textbf {\bibinfo {volume} {76}},\ \bibinfo {eid} {075301} (\bibinfo
  {year} {2007})},\ \Eprint {http://arxiv.org/abs/cond-mat/0702199}
  {arXiv:cond-mat/0702199 [cond-mat.mes-hall]} \BibitemShut {NoStop}%
\bibitem [{\citenamefont {Gruzberg}\ \emph {et~al.}(1999)\citenamefont
  {Gruzberg}, \citenamefont {Ludwig},\ and\ \citenamefont
  {Read}}]{GruzbergLudwigReadPRL1999}%
  \BibitemOpen
  \bibfield  {author} {\bibinfo {author} {\bibfnamefont {I.~A.}\ \bibnamefont
  {Gruzberg}}, \bibinfo {author} {\bibfnamefont {A.~W.~W.}\ \bibnamefont
  {Ludwig}}, \ and\ \bibinfo {author} {\bibfnamefont {N.}~\bibnamefont
  {Read}},\ }\href {\doibase 10.1103/PhysRevLett.82.4524} {\bibfield  {journal}
  {\bibinfo  {journal} {Phys. Rev. Lett.}\ }\textbf {\bibinfo {volume} {82}},\
  \bibinfo {pages} {4524} (\bibinfo {year} {1999})}\BibitemShut {NoStop}%
\end{thebibliography}%


%

\end{document}